\documentclass[12pt]{article}
\usepackage{amsfonts}
\usepackage{epsfig}
\usepackage{psfig}
\input{epsf}

\DeclareMathSymbol{\lesssim}      {\mathrel}{AMSa}{"2E}  
\DeclareMathSymbol{\gtrsim}       {\mathrel}{AMSa}{"26}

\def\3{\ss}                                           
\def\jpg#1#2#3  {{ J. Phys. G.} {#1} (#2) #3}
\def\cpc#1#2#3  {{ Comp. Phys. Comm.} {#1} (#2) #3}
\def\pl#1#2#3   {{ Phys. Lett. B} {#1} (#2) #3}
\def\EP#1#2#3   {{ Eur.  Phys. J. C} {#1} (#2) #3}
\def\prep#1#2#3 {{ Phys. Rep.} {#1} (#2) #3}
\def\prev#1#2#3 {{ Phys. Rev. D} {#1} (#2) #3}
\def\prl#1#2#3  {{ Phys. Rev. Lett.} {#1} (#2) #3}
\def\zp#1#2#3   {{ Z. Phys. C} {#1} (#2) #3}
\def\nim#1#2#3  {{ Nucl. Instr. and Meth. A} {#1} (#2) #3}
\def\np#1#2#3   {{ Nucl. Phys. B} {#1} (#2) #3}
\def\SET2{E^{2}_T}
\def\AVET1{\bar{E}_T}
\def\AVET2{\bar{E}^{2}_T}
\def\kt{k_T}

\def\Q2{Q^2}
\def\xbj{x}
\def\xgx{xg(x)}
\def\ETBO{E^{\rm BRE}_{T,1}}
\def\ETBT{E^{\rm BRE}_{T,2}}
\def\ETLO{E^{\rm LAB}_{T,1}}
\def\ETLT{E^{\rm LAB}_{T,2}}

\begin{document}
\vspace{+1cm}
\title {\bf\LARGE Dijet production  
                  in neutral current deep inelastic scattering at HERA}

\author{ZEUS Collaboration}

\date{  }

\maketitle

\vspace{-7.5cm}
\begin{flushleft}
\tt DESY 01-127 \\
August 2001 \\
\end{flushleft}
\vspace{+8cm}

\begin{abstract}

\noindent 
Dijet cross sections in neutral current deep inelastic $ep$ scattering have been 
measured in the range 10 $< \Q2 <$ 10$^4$ GeV$^2$ with the ZEUS detector at HERA using an 
integrated luminosity of 38.4 pb$^{-1}$. The cross sections, measured in the Breit 
frame using the $\kt$ jet algorithm, are compared with next-to-leading-order perturbative QCD 
calculations using proton parton distribution functions. The uncertainties of the QCD calculations 
have been studied. The predictions are in reasonable agreement with the measured cross sections 
over the entire kinematic range.

\end{abstract}
\newpage
\include{auth88_out}
\pagenumbering{arabic}
\newpage
\section{Introduction}

Dijet production in deep inelastic $ep$ scattering (DIS) provides a test of perturbative Quantum 
Chromodynamics (pQCD) and is sensitive to the structure of the proton. A comparison of measurements of 
dijet cross sections and pQCD predictions tests the validity of the concept of 
factorisation~\cite{factorisation} into a hard partonic cross section and universal parton distribution 
functions (PDFs). The large centre-of-mass energy of the HERA $ep$ collider 
($\sqrt{s} \approx $ 300 GeV) permits measurements of cross sections~\cite{prevpub,recent_h1}, for jets 
of high transverse energy, covering over three orders of magnitude in both the photon 
virtuality, $\Q2$, and the fraction of the proton's momentum carried by the struck parton, 
$\xbj$. 

\begin{figure}[htp]
  \epsfig{figure=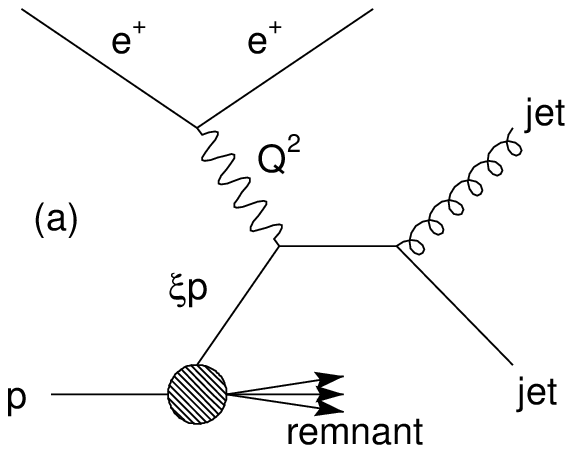,width=8.75cm}
  \epsfig{figure=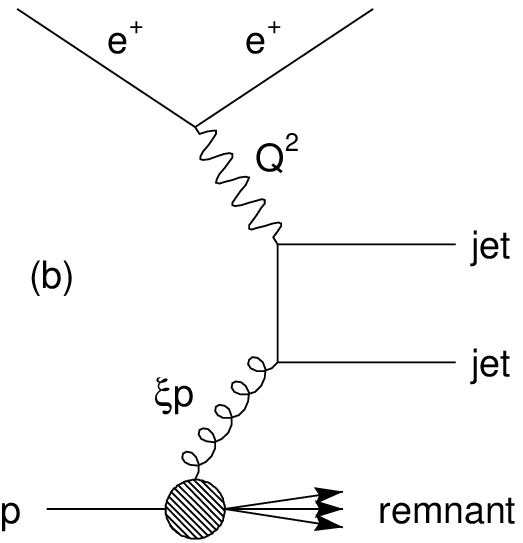,width=6.75cm}
  \caption{(a) The QCD-Compton diagram and 
    (b) the boson-gluon fusion diagram. 
    The fraction of the proton momentum, $p$, carried by the stuck
    parton is $\xi$.}
  \label{fig:F_diag}
\end{figure}

In leading-order (LO) QCD, two processes contribute to dijet production in DIS: the 
QCD-Compton (QCDC) and the boson-gluon fusion (BGF) processes, shown in 
Figs.~\ref{fig:F_diag}(a) and (b), respectively. The QCDC and the BGF cross sections are 
calculated in pQCD by convoluting the matrix element for the hard process, which depends 
upon the value of the QCD strong coupling constant, $\alpha_s$, with the PDFs of the proton. In the 
high-$\Q2$ region, the QCDC process is dominant and the quark PDFs are well constrained by inclusive 
DIS data. Hence the dijet cross-section measurements allow tests of pQCD and a measurement of $\alpha_s$. 
The extraction of $\alpha_s$ from the high-$\Q2$ dijet data and its evolution with $\Q2$ are the subjects 
of a separate publication~\cite{Etassi}. The BGF process is the dominant contribution to dijet production at 
$\Q2~\lesssim~500~{\rm GeV}^2$. Therefore, measurements of the dijet cross section at low 
$\Q2$ are sensitive to the gluon momentum distribution in the proton, $\xgx$, at low 
$\xbj$. The dijet cross section can be compared to the predictions 
of NLO QCD calculated using various parametrisations of the proton PDFs. These comparisons
complement analyses that determine $\xgx$ from the scaling violations 
of the structure functions~\cite{gluons_zeus,gluons_h1}. 

In this paper, the value of $\alpha_s (M_Z)$ is fixed and the inclusive dijet data ($\geq$ 2 jets) 
are used to test pQCD and the universality of the proton PDFs, in particular the gluon distribution.
The precision of the next-to-leading-order (NLO) QCD calculations plays an important 
role in this analysis and is also studied here in detail. The high-statistics dijet sample, 
together with improved NLO QCD calculations and a better understanding of jet-finding algorithms, 
permit higher-precision tests over a wider range of $\Q2$, $10~<~\Q2~<~10^4$~GeV$^2$, than in 
previous publications~\cite{prevpub}.

\section{Theoretical framework}

Within the framework of pQCD in DIS, the dijet production cross section, 
$d\sigma$, can be written as a convolution of the proton PDFs, $f_a$, 
with the partonic hard cross section, $d\hat{\sigma}_a$:
  
\begin{equation}
  d\sigma = \sum_{a=q,\bar{q},g}\int dx \ f_a(x,\mu_F^2) \ 
  d\hat{\sigma}_a (x,\alpha_s(\mu_R),\mu_R^2, \mu_F^2) 
  \cdot (1+\delta_{\rm had}).
\label{eq:xsec}
\end{equation}

The partonic cross section describes the short-distance structure of the interaction and 
is calculable as a power-series expansion in the strong coupling constant. As can be seen in 
Eq.~(\ref{eq:xsec}), the cross section depends on the renormalisation scale, $\mu_R$, since the 
calculations are not carried out to all orders. The PDFs contain the description of the long-distance 
structure of the incoming proton. The evolution of the PDFs with the factorisation scale, $\mu_F$, 
at which they are determined follows the DGLAP equations \cite{dglap}. The hadronisation 
correction, $\delta_{{\rm had}}$, can be estimated using Monte Carlo models 
for fragmentation (see Sections~\ref{sec:mc} and \ref{sec:had}).

The predictions of QCD were calculated at NLO in $\alpha_s$ using the programs 
MEPJET~\cite{MEPJET} and DISENT~\cite{DISENT}. These programs yield parton-level cross sections 
and allow for an arbitrary jet-definition scheme with user-defined cross-section cuts. Unless 
otherwise stated, all the pQCD predictions presented in this publication were 
calculated using the CTEQ4M~\cite{CTEQ} proton PDFs. The value, $\alpha_s(M_Z) = 0.116$, 
and the formula for the running of $\alpha_s$ used in MEPJET and DISENT were the same as 
those used in the proton PDF determination. According to calculations using DISENT, the percentage 
of gluon-initiated dijet events in the total dijet cross section increases from $\sim10\%$ to $\sim70\%$ 
as $\Q2$ decreases from $10^4$ to 10~GeV$^2$.

In all NLO QCD calculations, $\mu_F^2$ was set to $Q^2$. The dependence of the pQCD predictions 
on $\mu_F^2$ and $\mu_R^2$ is discussed in Section~\ref{sec:scale}. In DIS dijet events, there are two ``natural'' 
variables to define $\mu_R^2$: the $Q^2$ of the event and the square of the transverse energy  of 
the event or jets. In principle, any multiple of these variables or any combination of 
them can be used. The interplay between these two variables has been investigated  by measuring the 
dependence of the dijet cross section on the ratio $\AVET2/Q^2$, where $\bar{E}_T$ is the average 
transverse energy of the two jets  with highest transverse energy in the event.

\section{Experimental setup}

The dijet data sample presented here was collected with the ZEUS detector 
during 1996 and 1997 and corresponds to an integrated luminosity of 
$38.4\pm0.6{\ {\rm pb^{-1}}}$. During this period, HERA operated with 
protons of energy $E_p = 820$~GeV and positrons of energy $E_e = 27.5$~GeV. 

The ZEUS detector is described in detail elsewhere~\cite{sigtot,status}.
The main components used in the present analysis are the 
uranium-scintillator sampling calorimeter (CAL) \cite{main}, and the central 
tracking chamber (CTD)~\cite{CTD} positioned in a 1.43~T solenoidal magnetic 
field. The smallest subdivision of the CAL is called a cell. The CAL relative energy resolutions, as measured in test beams, are 
$18\%/\sqrt{E(\mbox{GeV})}$ and $35\%/\sqrt{E(\mbox{GeV})}$ for 
electrons and hadrons, respectively. The interaction vertex is measured 
using the CTD with a typical resolution along (transverse to) the beam 
direction\footnote{The 
ZEUS coordinate system is a right-handed Cartesian system, with the $Z$ axis 
pointing in the proton beam direction, referred to as the 
``forward direction'', and the $X$ axis pointing left towards the centre of 
HERA. The coordinate origin is at the nominal interaction point. The 
pseudorapidity is defined as $\eta=-\ln\left(\tan\frac{\theta}{2}\right)$, 
where the polar angle, $\theta$, is 
measured with respect to the proton beam direction.} of 0.4 (0.1) cm.

The luminosity was measured using the Bethe-Heitler reaction 
$e^+p \rightarrow e^+\gamma p$~\cite{LUMI}. The resulting small-angle 
energetic photons were measured by the luminosity monitor, a lead-scintillator 
calorimeter placed in the HERA tunnel at $Z = -107$ m.

\section{Reconstruction of kinematic variables}

The reaction
\[
e^+(k) + p(P) \rightarrow e^+(k') + X 
\]
at fixed squared centre-of-mass energy, $s=(k+P)^2$, can be fully specified in terms of 
$\Q2 \equiv -q^2= -(k-k')^2$ and Bjorken $\xbj = \Q2/(2P\cdot q)$. The fraction of the 
positron's energy transferred to the proton in its rest frame is $y = \Q2/(s\xbj)$. 

For processes where two or more jets are produced in the final state, the observable 
$\xi$ is defined as

\[
  \xi = \xbj \ \left(1+\frac{M_{JJ}^2}{\Q2}\right),
\]

where the dijet mass, $M_{JJ} = \sqrt{2 E_1  E_2 ( 1 - \cos\theta_{12})}$, is calculated from 
the energies $E_1$ and $E_2$ and the opening angle $\theta_{12}$ of the two jets of 
highest transverse energy. The variable $\xi$ is the fraction of the proton momentum 
carried by the struck parton in the LO QCD processes (see Fig.~\ref{fig:F_diag}).

The kinematic variables were reconstructed using a combination of the electron 
and the double-angle (DA) methods~\cite{DA}. The electron method was used 
except when the angle of the hadronic system, $\gamma_h$, 
was less than 90$^\circ$ and the scattered-positron track could be 
well reconstructed by the CTD. In this case, the DA method was used. 

The variable $y_{\rm JB} = \sum_i E_i (1-\cos \theta_i) /(2 E_e)$, calculated 
according to the Jacquet-Blondel method~\cite{YJB}, where 
the sum runs over all CAL cells except those belonging to the 
scattered positron, gives a measurement of $y$ with good resolution at 
low $y$. 

\section{Selection of the dijet event sample}

The events were selected online via a three-level trigger system~\cite{status,zeuscc94,zeushix} 
using the same selection algorithms as in previous dijet publications~\cite{virtual}. Neutral 
current DIS events were selected by requiring that the scattered positron was measured in the 
CAL~\cite{ZNCpaper}. Further criteria were applied both to ensure an accurate reconstruction 
of the kinematic variables and to increase the purity of the sample:

\begin{itemize}

\item $E_e^\prime >10$ GeV, where $E_e^\prime$ is the scattered positron 
energy, after   correction for energy loss in inactive material in front 
of the CAL, to achieve a high-purity sample of DIS events;

\item $38 \le E-p_Z \le 65$~GeV, where $E-p_Z = \sum_i E_i (1-\cos \theta_i)$ 
and the summation is over all CAL cells, to remove background from photoproduction and events with 
large initial-state QED radiation;

\item $y_e\leq 0.95$, where $y_e=1-\frac{E_e^\prime}{2E_e}(1-\cos\theta_e^\prime)$ and 
$\theta_e^\prime$ is the polar angle of the scattered positron. Along with the previous 
requirements, this reduces the photoproduction background to a negligible level~\cite{f2};

\item $ |X|  >  14\ {\rm cm}\ {\rm or}\ |Y|  >  14\ {\rm cm} ,$ where 
$X$ and $Y$ are the impact positions of the positron on the CAL, to 
avoid the low-acceptance region adjacent to the rear beampipe;

\item $|Z_{\rm vertex}|  <  50\ {\rm cm}$, to ensure that event quantities 
can be   accurately determined;

\item $y_{JB}\geq 0.04$, to give sufficient accuracy for DA reconstruction 
of $Q^2$  and $x$;

\item 10 $< Q^2 <$ 10$^4$ GeV$^2$.

\end{itemize}

After these cuts, jets  were reconstructed using the longitudinally invariant $\kt$ cluster 
algorithm~\cite{kt} in the inclusive mode. The jet search was conducted in the Breit 
frame~\cite{BREIT}, defined by ${\bf q} + 2 \xbj {\bf P} = 0$, where ${\bf q}$ and ${\bf P}$ are 
the three-momentum vectors of the exchanged boson and the proton, respectively.
In the Breit frame, the three-momentum vectors of the exchanged boson and  
the incoming and outgoing partons all lie on the $Z$ axis for a quark-parton-model 
type of event, i.e. the production of a single quark in the final state. For a typical 
QCDC or BGF event, the transverse energy in this frame is non-zero. The vector required to boost 
to the Breit frame was determined using the reconstructed event kinematics. 

For each event, the jet search was 
performed over all the CAL energy deposits, considered as massless objects
and boosted to the Breit frame, excluding those corresponding to the 
identified scattered positron candidate. The jet transverse energies were  
corrected for energy loss in the inactive material in front of the CAL. 
Events with two or more jets found in the Breit frame were selected by requiring that 
the two jets with the highest transverse energy satisfied the following cuts:

\begin{itemize}

\item $\ETBO > 8$ GeV and $\ETBT > 5$ GeV, where $\ETBO$ and $\ETBT$ are the 
      transverse energies in the Breit frame;  

\item $\ETLO > 5$ GeV and $\ETLT > 5$ GeV, where $\ETLO$ and $\ETLT$ are the 
      transverse energies of the two jets in the laboratory frame; 

\item $|\eta^{\rm LAB}_{1,2}| < 2$, where $\eta^{\rm LAB}_{1,2}$ are the pseudorapidities 
      of the two jets in the laboratory frame.

\end{itemize} 

The jet transverse energy and pseudorapidity cuts in the laboratory frame were imposed in order to select 
well-measured jets within the acceptance of the CAL. After all cuts, 39576 events with two or more jets,
including 3902 events with three or more jets, remained in the sample. 

\section{Monte Carlo simulation}
\label{sec:mc}

Monte Carlo (MC) simulations were used to correct the data for detector acceptance and resolution. 
Two MC models were used to generate DIS events: ARIADNE~4.10~\cite{ariadne} 
and 
LEPTO~6.5~\cite{LEPTO}. In ARIADNE, the QCD cascade is simulated using the colour-dipole model. 
LEPTO uses the exact matrix elements to generate the hard process and the parton-shower model to 
simulate higher-order processes. Both models use the Lund string-fragmentation model~\cite{string} 
for hadronisation, as implemented in JETSET~7.4 \cite{JETSET}. To take into account first-order 
electroweak corrections, LEPTO and ARIADNE were interfaced with HERACLES~4.5.2~\cite{HERACLES} 
using the DJANGO6~2.4 \cite{DJANGO} program. The CTEQ4D proton PDFs were used. To estimate the 
uncertainty due to hadronisation, events were also produced using the HERWIG~5.9 
generator~\cite{HERWIG}, in which the fragmentation is simulated using a cluster model~\cite{CLUSTER}.

The ZEUS detector response was simulated with a program based on 
GEANT~3.13~\cite{GEANT}. The generated events were passed through the simulated detector, 
subjected to the same trigger requirements as the data and processed by the same 
reconstruction and offline programs.

\begin{figure}[htp]
\begin{center}
\epsfig{figure=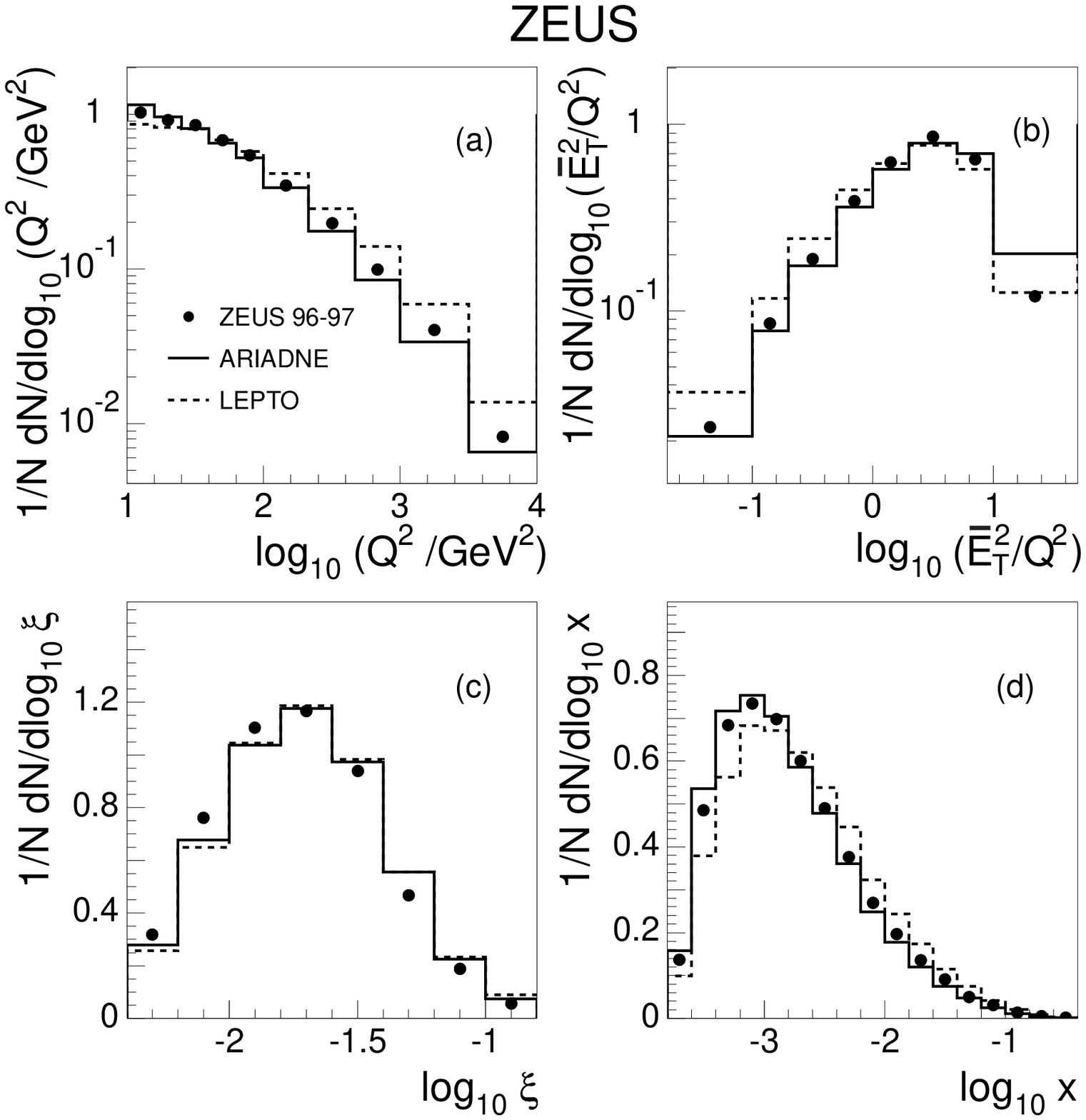,width=16.0cm}
\caption{ Normalised uncorrected distributions of
  (a) log$_{10}(Q^2/{\rm GeV}^2)$, (b) log$_{10}(\AVET2/Q^2)$, 
(c) log$_{10}\xi$ and 
  (d) log$_{10}\xbj$ for dijet events selected with the $\kt$ algorithm in the
  Breit frame, for $E^{\rm BRE}_{T,1} >$ 8 GeV, $E^{\rm BRE}_{T,2} >$
  5 GeV, $E^{\rm LAB}_{T,1} >$ 5 GeV, $E^{\rm LAB}_{T,2} >$ 5 GeV,
  $|\eta^{\rm LAB}_{1,2}|~<$~2, $y>$ 0.04 and E$_e^{'} >$ 10 GeV in the 
  range 10 $<Q^2$ $<$ 10$^4$ GeV$^2$.  The points are the measurements. The statistical 
  uncertainties are generally smaller than the symbols. 
  The solid and dashed histograms are the predictions of the ARIADNE~4.10
  and LEPTO~6.5 Monte Carlo programs, respectively.}
\label{fig:DATA_MC1}
\end{center}
\end{figure}

\begin{figure}[htp]
\begin{center}
\epsfig{figure=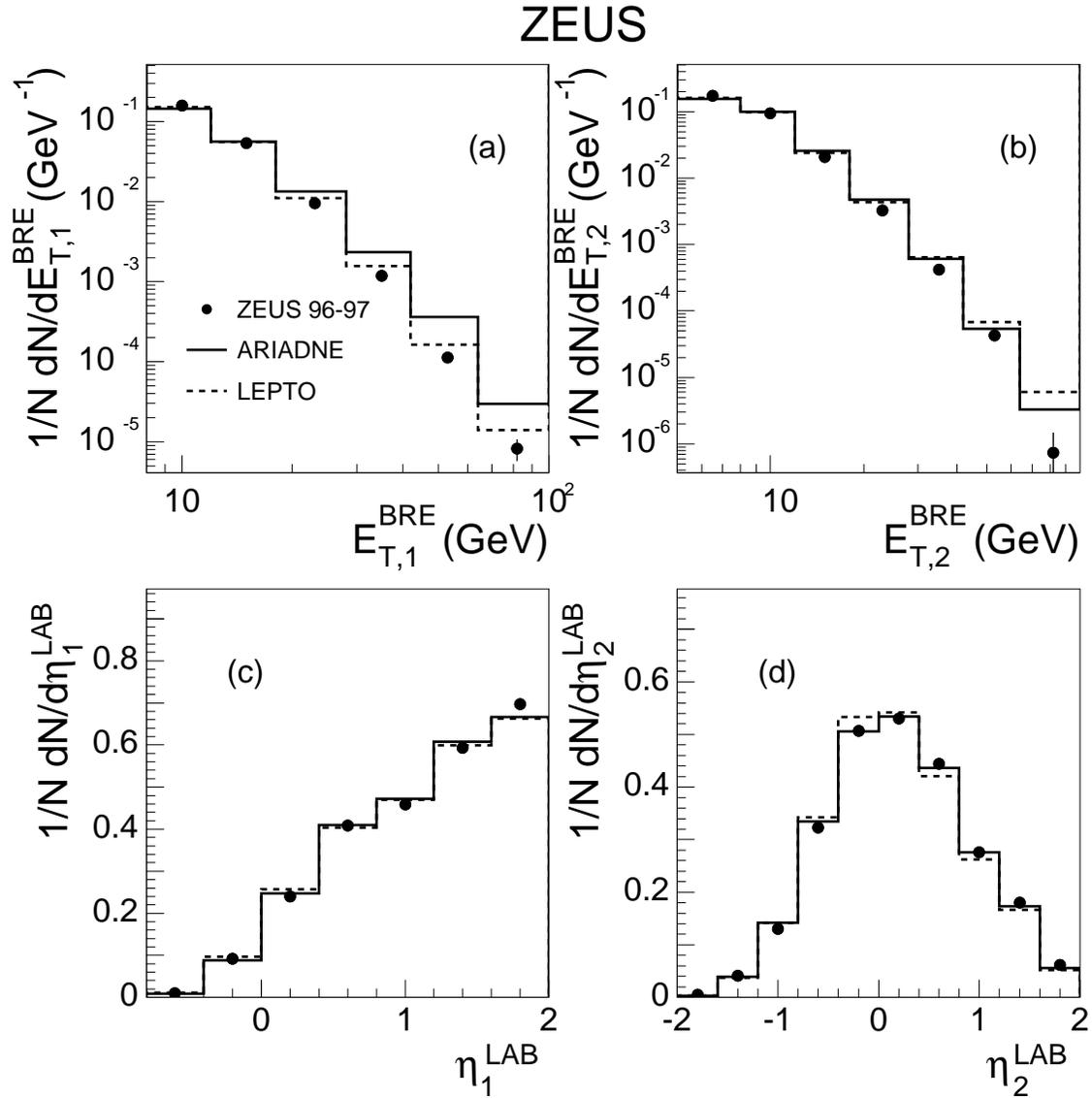,width=16.0cm}
\caption{Normalised uncorrected distributions of (a) $E^{\rm BRE}_{T,1}$, 
  (b) $E^{\rm BRE}_{T,2}$, (c) $\eta_1^{\rm LAB}$ and (d) $\eta_2^{\rm LAB}$ for dijet
  events selected with the $\kt$ algorithm in the Breit frame.
  For details, see the caption of Fig. 2.}
\label{fig:DATA_MC2}
\end{center}
\end{figure}

Figures~\ref{fig:DATA_MC1} and~\ref{fig:DATA_MC2} show normalised uncorrected differential 
distributions of  dijet events together with the predictions of the ARIADNE and LEPTO MC 
programs. As shown in Fig.~\ref{fig:DATA_MC1}(a), (b) and (d), the global event variables are 
better described by ARIADNE, while both ARIADNE and LEPTO give an adequate 
description of the data in Fig.~\ref{fig:DATA_MC1}(c). For the jet quantities, 
LEPTO gives a better description of the data in Fig.~\ref{fig:DATA_MC2}(a), while for the three 
other distributions, ARIADNE is better overall. Therefore, ARIADNE was used as the 
default MC simulation to determine the corrections from the detector to the hadron level.

\section{Theoretical uncertainties}

\subsection{Uncertainties due to the renormalisation and factorisation scales}
\label{sec:scale}

A study of the uncertainties introduced in the theoretical predictions from the renormalisation 
scale has been performed by choosing $\mu_R^2$ to be either $\Q2$ or $E_T^2/4$, where $E_T^2$ 
is the square of the sum of the transverse energies of all final-state partons, and by varying 
each by a factor of four. 

\begin{figure}[htp]
\begin{center}
\epsfig{figure=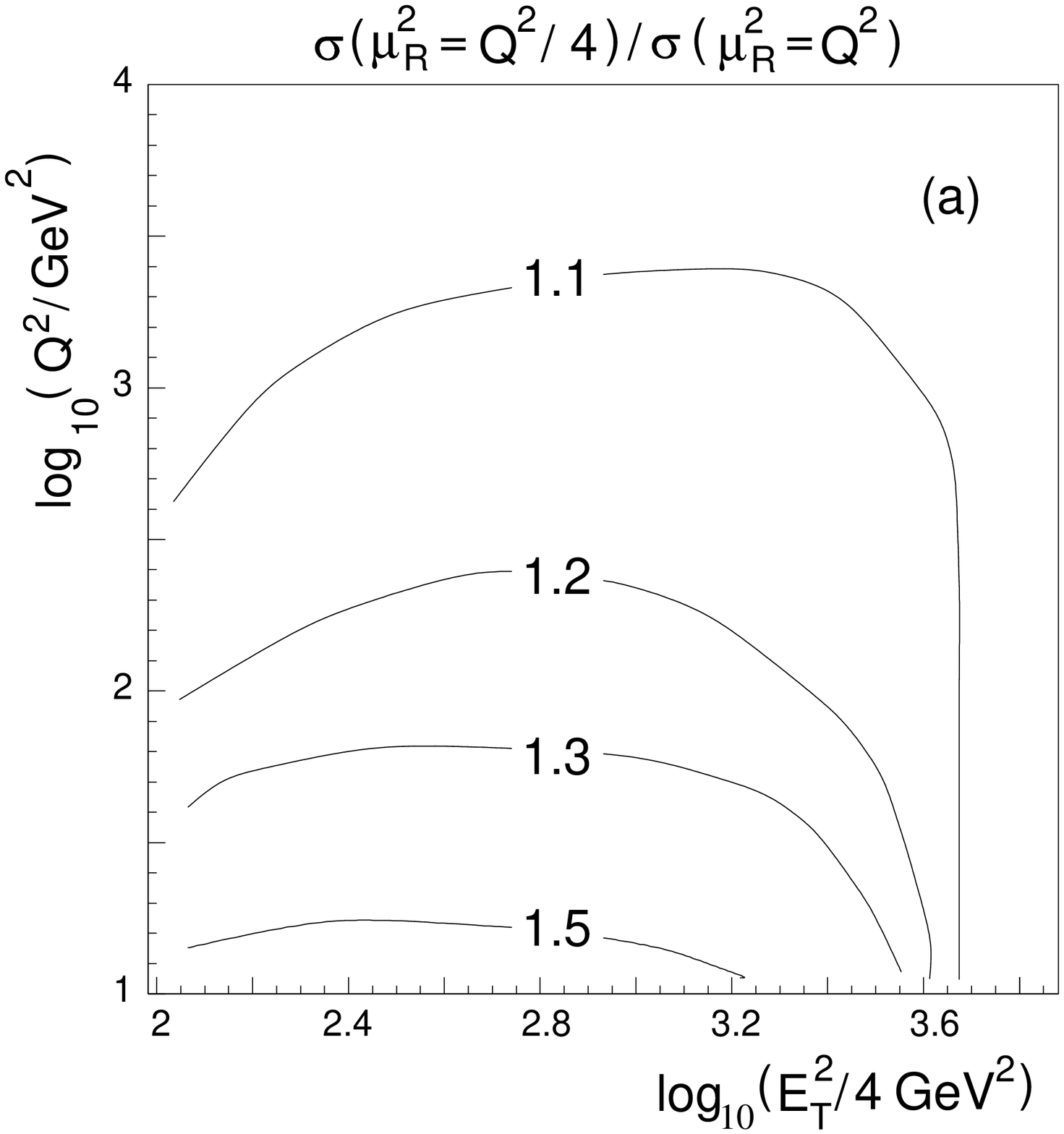,height=8.5cm}
\epsfig{figure=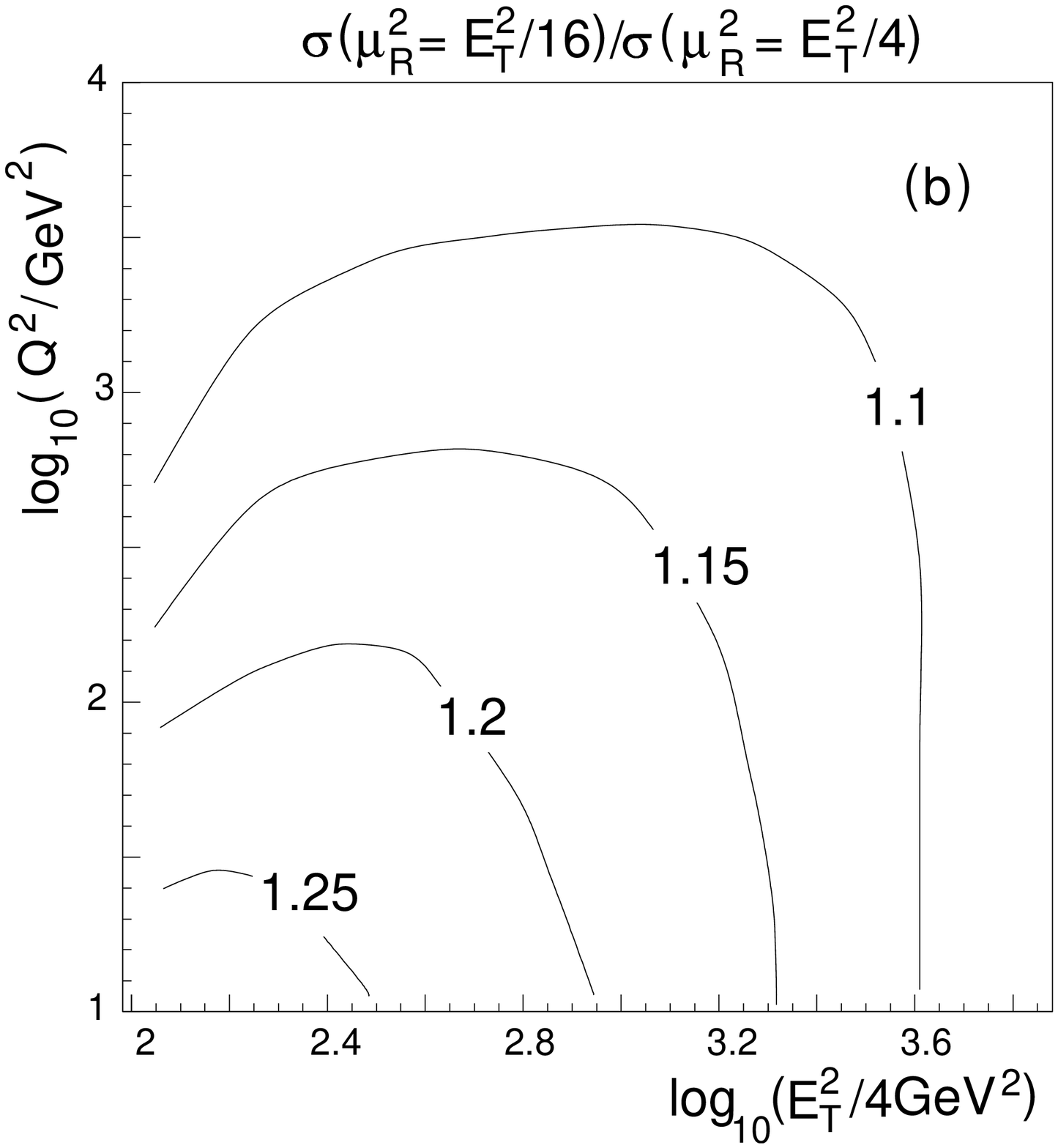,height=8.5cm}
\end{center}
\caption{
  Contours representing a fixed scale uncertainty of the NLO QCD calculations as a function of 
  $\log_{10}\Q2$ and $\log_{10}(\SET2/4)$, where $\Q2$ and $E_T^2$ are in GeV$^2$.
  The scale uncertainty shown in (a) was calculated by taking the
  ratio of the predicted dijet cross sections
  $\sigma(\mu_R^2=\Q2/4)/\sigma(\mu_R^2=\Q2)$; that shown in (b)
 was calculated by taking the ratio of the predicted dijet cross
  sections $\sigma(\mu_R^2=\SET2/16)/\sigma(\mu_R^2=\SET2/4)$.  The
  cross sections were calculated using DISENT and refer to jets
  of partons selected with the inclusive $\kt$ jet algorithm in the
  Breit frame. CTEQ4M was used for the proton PDFs.}
\label{fig:scale_q2_et}
\end{figure}

The ratio of the predicted dijet cross sections, $\sigma(\mu_R^2= \Q2/4)/\sigma(\mu_R^2=\Q2)$, as 
a function of $\Q2$ and $\SET2/4$ calculated using DISENT is shown in Fig.~\ref{fig:scale_q2_et}(a). 
The theoretical uncertainty depends mainly on $\Q2$ and decreases from values of about 50$\%$ at 
low $\Q2$ to about 10$\%$ at high $\Q2$. The ratio of the predicted dijet cross sections, 
$\sigma(\mu_R^2= \SET2/16)/\sigma(\mu_R^2=\SET2/4)$, as a function of $\Q2$ and $\SET2/4$ calculated 
using DISENT is shown in Fig.~\ref{fig:scale_q2_et}(b). The estimated theoretical uncertainty 
observed in Fig.~\ref{fig:scale_q2_et}(b) is always less than that shown in 
Fig.~\ref{fig:scale_q2_et}(a) and has a similar dependence on both $\Q2$ and $\SET2$/4 over the entire 
kinematic range. Results similar to those shown in Fig.~\ref{fig:scale_q2_et} were obtained when 
calculating the ratios $\sigma(\mu_R^2= 4\Q2)/\sigma(\mu_R^2=\Q2)$ and 
$\sigma(\mu_R^2= \SET2)/\sigma(\mu_R^2=\SET2/4)$. 
Even for $E_T$ as high as 100~GeV, the scale uncertainty of the pQCD calculations 
is smaller than $10\%$ only for $\Q2$ greater than 2000~GeV$^2$.
     
The uncertainties introduced in the theoretical predictions from the factorisation scale were studied
by changing the value of $\mu_F^2=Q^2$ by a factor of four. The resulting change in the 
cross section was about 5$\%$, significantly less than when $\mu_R^2$
was varied by the same factor.

\subsection{Hadronisation uncertainty}
\label{sec:had}

The results of the NLO QCD programs for calculating dijet production are given in terms of jets of 
partons, while the measured cross sections refer to jets of hadrons. In both 
cases, the jets were 
reconstructed using the longitudinally invariant $k_T$ algorithm in its inclusive mode. Therefore, 
a correction was applied to the predicted parton-level cross sections to account for the effects of 
hadronisation. Correction factors were defined for each bin as the ratio of the parton-level to 
hadron-level cross sections and were calculated using the MC simulations. The predictions of NLO QCD 
were divided by the mean of the  correction factors calculated using the ARIADNE and LEPTO MC programs. 
Since LEPTO with default parameters does not agree with the data as well as ARIADNE (see 
Fig.~\ref{fig:DATA_MC1}), the LEPTO sample was reweighted to agree with the $\Q2$ distribution of the 
data. After this reweighting, all other kinematic distributions were found to agree better with the 
data. The difference in the LEPTO correction factors before and after reweighting was negligible. 

The correction factors as a function of $\Q2$ were found to be $\sim$1.25 at low $\Q2$, falling to 
$\sim$1.15 for $\Q2 >$ 100 GeV$^2$. As a function of $\log_{10}\xi$ and $\AVET2/Q^2$, the correction 
was in the range $1.15-1.30$. The difference between the mean correction factor and the individual 
correction factors from LEPTO and ARIADNE was defined as the uncertainty of the theoretical 
predictions due to uncertainties in the parton-to-hadron corrections. On average, this difference is about 5$\%$. 
It is largest ($12\%$) in the range $-2.4 < \log_{10}\xi < -2.2$, while, for $\Q2 <$ 25 GeV$^2$, a 
difference of about $10\%$ was observed. The correction factors obtained from HERWIG, which uses a 
different hadronisation model than either LEPTO or ARIADNE, were also calculated and were found to 
deviate by typically less than 5$\%$ from the correction factors obtained with ARIADNE.

\section{Data corrections and systematic uncertainties}

The cross sections for jets of 
hadrons in bins of $\log_{10}Q^2$, $\log_{10}\xi$ and $\log_{10}(\AVET2/Q^2)$ were obtained by applying a bin-by-bin correction to the measured dijet distributions using 
ARIADNE. The corrections take into account the efficiency of the trigger, the selection criteria and 
the purity and efficiency of the jet reconstruction. An additional MC correction was applied to the 
measured cross sections to account for initial- and final-state QED-radiation effects. 

A detailed study of the systematic uncertainties of the measurements was 
performed~\cite{mp_thesis,dc_thesis}. The uncertainty due to that of the jet-energy scale was studied 
by selecting a separate sample of DIS events with only one jet in the final state in the laboratory 
frame. In this sample, the transverse energy of the scattered positron, which is known to within 
$\pm 1\%$~\cite{ZNCpaper}, balanced the transverse energy of the jet. The ratio between the jet and 
the positron transverse energies was evaluated as a function of the jet pseudorapidity, separately 
for data and MC-generated events. A difference of up to $\pm 3\%$ was observed, resulting in a 
systematic uncertainty of about $\pm 10\%$ for the measured cross sections throughout the kinematic 
range. The jet-energy scale dominates all other uncertainties except at $\Q2~>$~1000~GeV$^2$, where 
the statistical plus other systematic uncertainties become comparable.

The other systematic uncertainties originate from the residual uncertainties in the event simulation 
and were estimated by correcting the data using the LEPTO correction factors and by varying the cuts 
in both data and MC simulation by an amount equal to the resolution on the relevant quantity. The most 
important were:

\begin{itemize}

\item {\it use of the MC model} - using LEPTO  instead of ARIADNE to evaluate the acceptance corrections 
      resulted in an uncertainty of typically $\pm5\%$;

\item {\it jet-transverse-energy cuts} - $\ETBO$, $\ETBT$, $\ETLO$ and $\ETLT$ were simultaneously varied 
      by the jet-transverse-energy resolution near the cuts, $\pm 17\%$, to account for differences 
      between the data and the MC simulation. This resulted in an uncertainty of up to $\pm12\%$ and 
      typically of $\pm4\%$;

\item {\it jet-pseudorapidity cuts} - a change of $\pm$0.1 in the $\eta^{\rm LAB}_{1,2}$ cuts imposed on 
      the jets in the laboratory frame resulted in an uncertainty of up to $\pm6\%$ and typically of 
      $\pm2\%$;

\item {\it $E-P_Z$ cuts} - the lower $E-P_Z$ cut of 38 GeV was varied by $\pm 4$ GeV, resulting in an 
      uncertainty of less than $\pm2\%$.

\end{itemize}

All the uncertainties are correlated between bins except for that due to the acceptance correction using 
LEPTO. Nevertheless, they have been added in quadrature when displayed in the figures and tables, with 
only the uncertainty due to the jet-energy scale shown separately.

\section{Results and discussion}

\subsection{Asymmetric jet cuts for the dijet cross section}

\begin{figure}[htp]
\begin{center}
\epsfig{figure=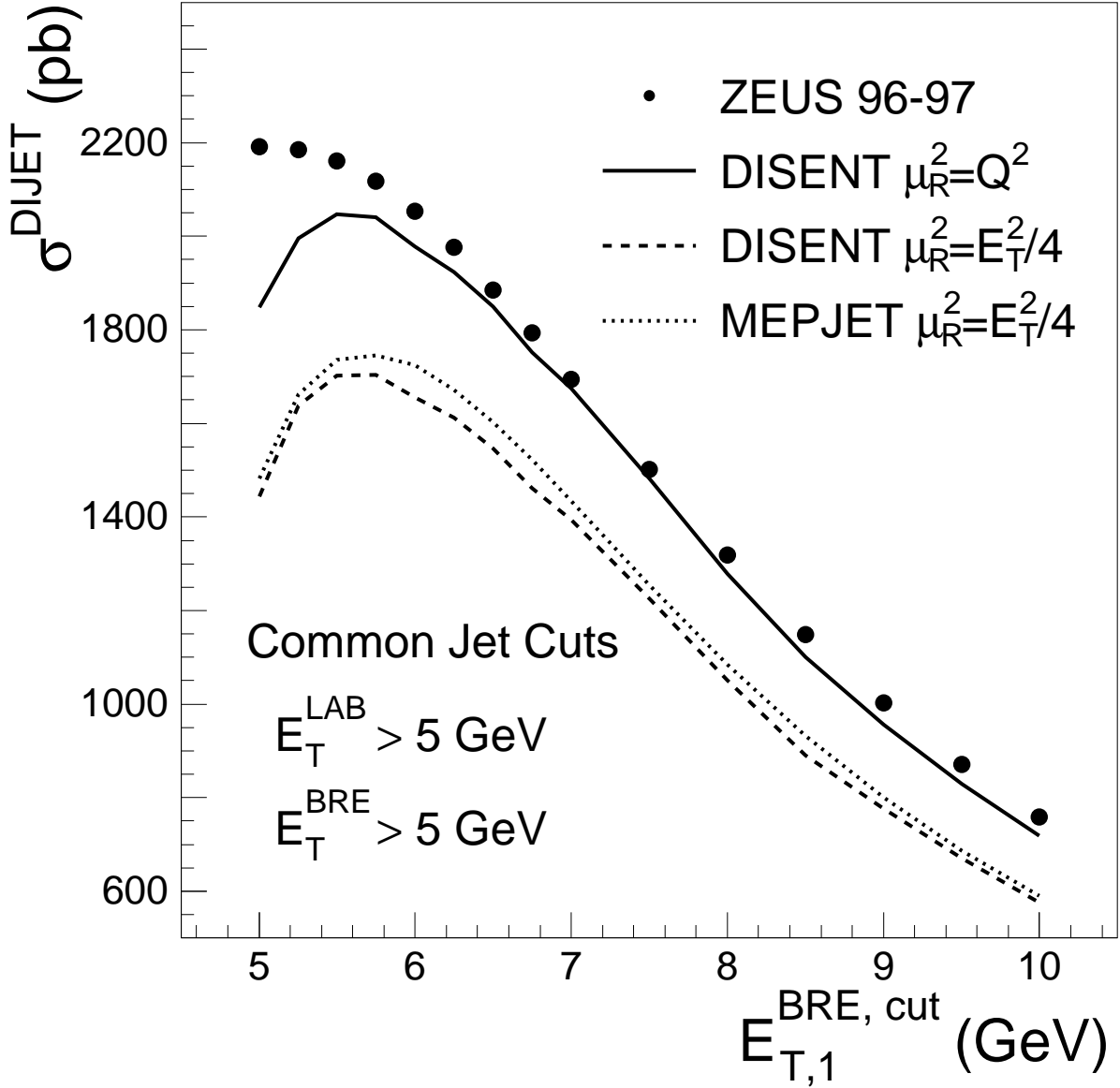,width=16.0cm}
\caption{The dijet cross section for jets of hadrons in the Breit frame,
  selected with the inclusive $\kt$ algorithm as a function of $E_{T,1}^{\rm BRE, cut}$.  
  The points represent the measured cross section. The typical magnitude
  of the statistical and systematic uncertainties added in quadrature
  is $\pm 15\%$. The full line represents the predictions of NLO QCD using DISENT
  with $\mu_R^2=Q^2$. The dashed line is the prediction of DISENT with
  $\mu_R^2=E_T^2/4$. The dotted line is the prediction of MEPJET with
  $\mu_R^2=E_T^2/4$. All three pQCD calculations use the CTEQ4M proton 
  PDFs, refer to jets of
  hadrons and were obtained by dividing the parton-level predictions
  of DISENT by correction factors computed using the ARIADNE~4.10
  MC simulation.}
\label{fig:cut_study}
\end{center}
\end{figure}

Figure~\ref{fig:cut_study} shows the measured and predicted inclusive dijet cross sections at the 
hadron level as a function of the threshold on the transverse
energy of the leading jet, $E_{T,1}^{\rm BRE, cut}$. 
For this study, the requirement on the jet with the highest transverse energy was relaxed and the phase space 
region was defined by: 10 $< Q^2 <$ 10$^4$ GeV$^2$, $y > $ 0.04, $E^\prime_e> 10$~GeV, $\ETBO~>~5$~GeV, 
$\ETBT > 5$ GeV, $\ETLO > 5$ GeV, $\ETLT > 5$~GeV and $|\eta^{\rm LAB}_{1,2}| < 2$. The NLO QCD 
predictions of DISENT with $\mu_R^2 = \Q2$ exhibit an unphysical behaviour at low $E_{T,1}^{\rm BRE, cut}$; 
as the $\ETBO$ threshold decreases below 6 GeV, the predicted cross section decreases, whereas the dijet cross 
section should increase. This occurs because, within NLO QCD near the ``symmetric-cut'' threshold at 
which $E_{T,1}^{\rm BRE, cut} = E_{T,2}^{\rm BRE, cut}$, the phase space for the real emission of a 
third parton is reduced. This results in an incomplete cancellation between the real emission and the 
virtual-loop corrections~\cite{Potter} and hence the cross section falls. As seen in 
Fig.~\ref{fig:cut_study}, the predictions of DISENT with $\mu_R^2 = \Q2$ approach the measured cross 
section for $E_{T,1}^{\rm BRE, cut} \gtrsim$ 6.5 GeV. 

To make comparisons of the measured dijet cross sections with the NLO QCD predictions, an asymmetric 
cut on the jet transverse energies of $\ETBO > $ 8 GeV and $\ETBT > $ 5 GeV was used for all further 
analysis. As seen in Fig.~\ref{fig:cut_study}, $E_{T,1}^{\rm BRE, cut}$ $=$  8 GeV is sufficiently far 
away from the unphysical region near the symmetric-cut threshold. The predictions of NLO QCD with this 
cut, using DISENT with $\mu_R^2 = \Q2$, agree with the data to within 5$\%$. This 5$\%$ difference between 
the predictions and the data is well within the estimated $\pm 15\%$ systematic uncertainty of the 
measurements.

The NLO QCD predictions using DISENT are also shown for the choice of $\mu_R^2 = E_T^2/4$ and underestimate 
the measurements by $20\%$. Calculations were also done using MEPJET. The predictions of MEPJET agree with 
those of DISENT to within 5$\%$.

\subsection{Measurements of the dijet cross sections}

\begin{figure}[htp]
\begin{center}
\epsfig{figure=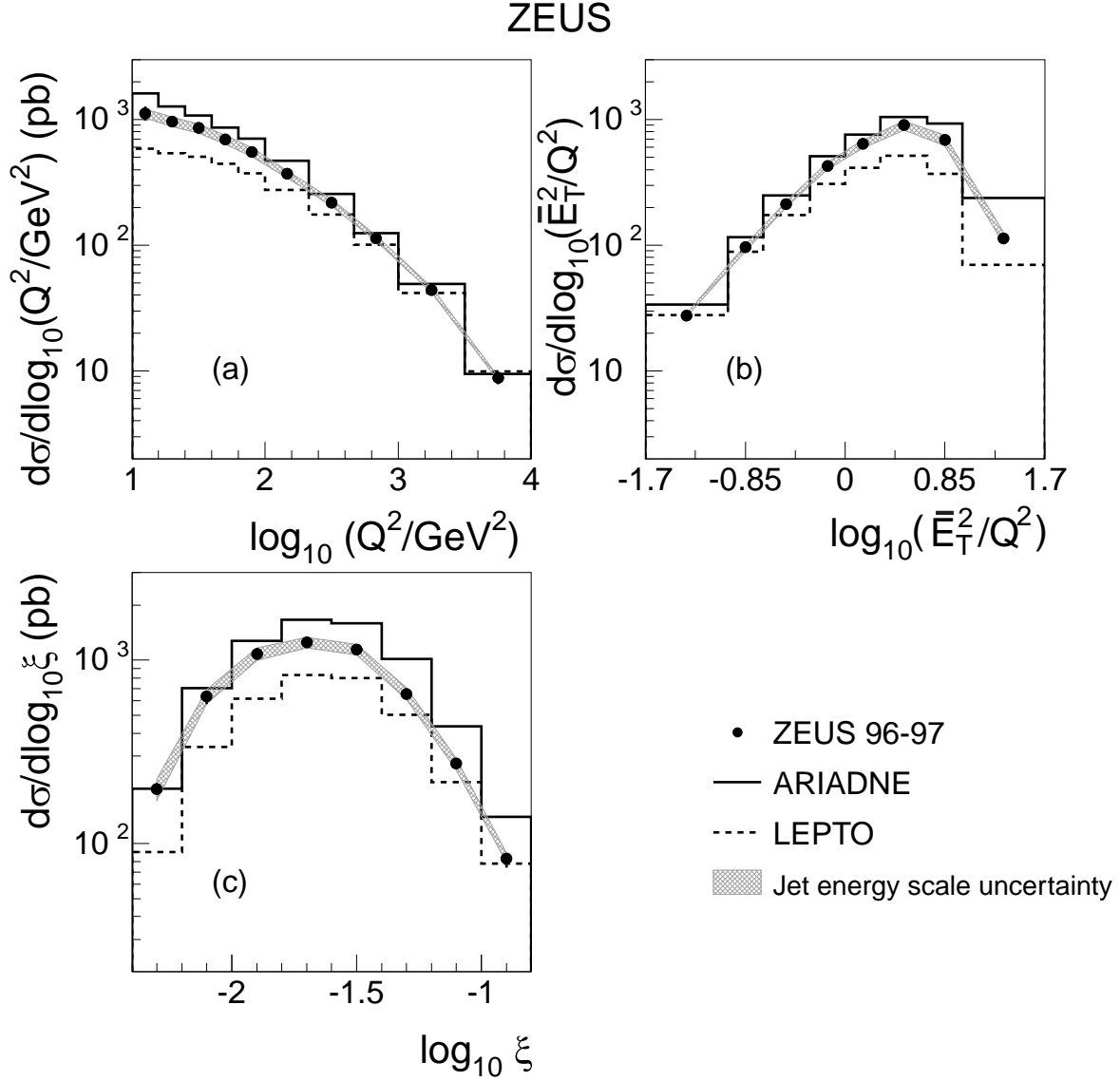,width=16.0cm}   
\caption{Dijet cross sections (a) $d\sigma/d\log_{10}(Q^2/{\rm GeV}^2)$,
  (b) $d\sigma/d\log_{10}(\AVET2/\Q2)$ and (c) $d\sigma/d\log_{10}\xi$ 
  for jets of hadrons in the Breit frame selected with the inclusive $\kt$ algorithm.
  The points represent the measured cross sections.  The error bars 
  are generally smaller than the points. The shaded band represents the systematic uncertainty
  due to the jet-energy scale.  The full (dashed) histogram represents
  the predictions of ARIADNE~4.10 (LEPTO~6.5) with the CTEQ4M proton PDFs.}
\label{fig:had_sigmas}
\end{center}
\end{figure}

\begin{figure}[htp]
\begin{center}
\epsfig{figure=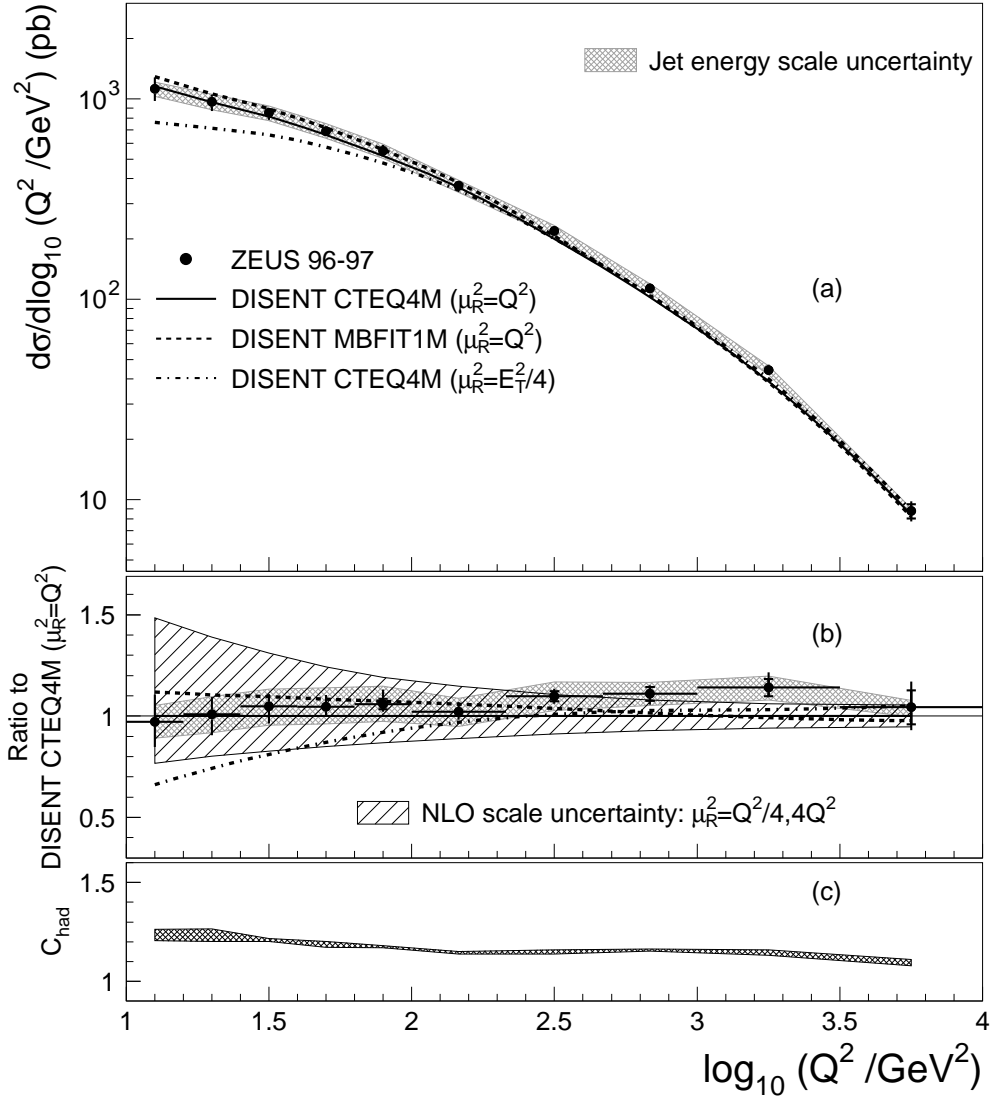,width=13.5cm}
\caption{(a) Dijet cross section, 
  $d\sigma/d\log_{10}\Q2$, for jets of hadrons 
  in the Breit frame selected with the
  inclusive $\kt$ algorithm.  The points represent the measured cross
  sections.  The inner error bars represent the statistical
  uncertainties and the outer bars are the statistical and the 
  systematic uncertainties added in quadrature.  The shaded band
  represents the systematic uncertainty due to the jet-energy scale.
  The full
  line represents the predictions of NLO QCD using DISENT
  with $\mu_R^2=Q^2$ and the CTEQ4M proton PDFs. The dashed-dotted line
  is the prediction of DISENT with $\mu_R^2=E_T^2/4$ and the CTEQ4M
  PDFs.  The dashed line is the prediction of DISENT with
  $\mu_R^2=Q^2$ using the MBFIT1M proton PDFs.  (b) The cross sections
  in (a) divided by the predictions of DISENT with $\mu_R^2=Q^2$ and
  the CTEQ4M PDFs.  The cross-hatched band represents the theoretical
  uncertainty due to the choice of $\mu^2_R$ calculated by choosing
  $\mu^2_R=\Q2/4$ and $\mu_R^2=\Q2$.  All of the pQCD
  predictions shown here refer to jets of hadrons and were
  obtained by dividing the parton-level predictions of DISENT by the
  average value of the correction factors computed using the LEPTO~6.5
  and the ARIADNE~4.10 MC simulations. The shaded band in (c) shows the magnitude
  and the uncertainty of this parton-to-hadron correction.}
\label{fig:q2_cross}
\end{center}
\end{figure}

\begin{figure}[htp]
\begin{center}
\epsfig{figure=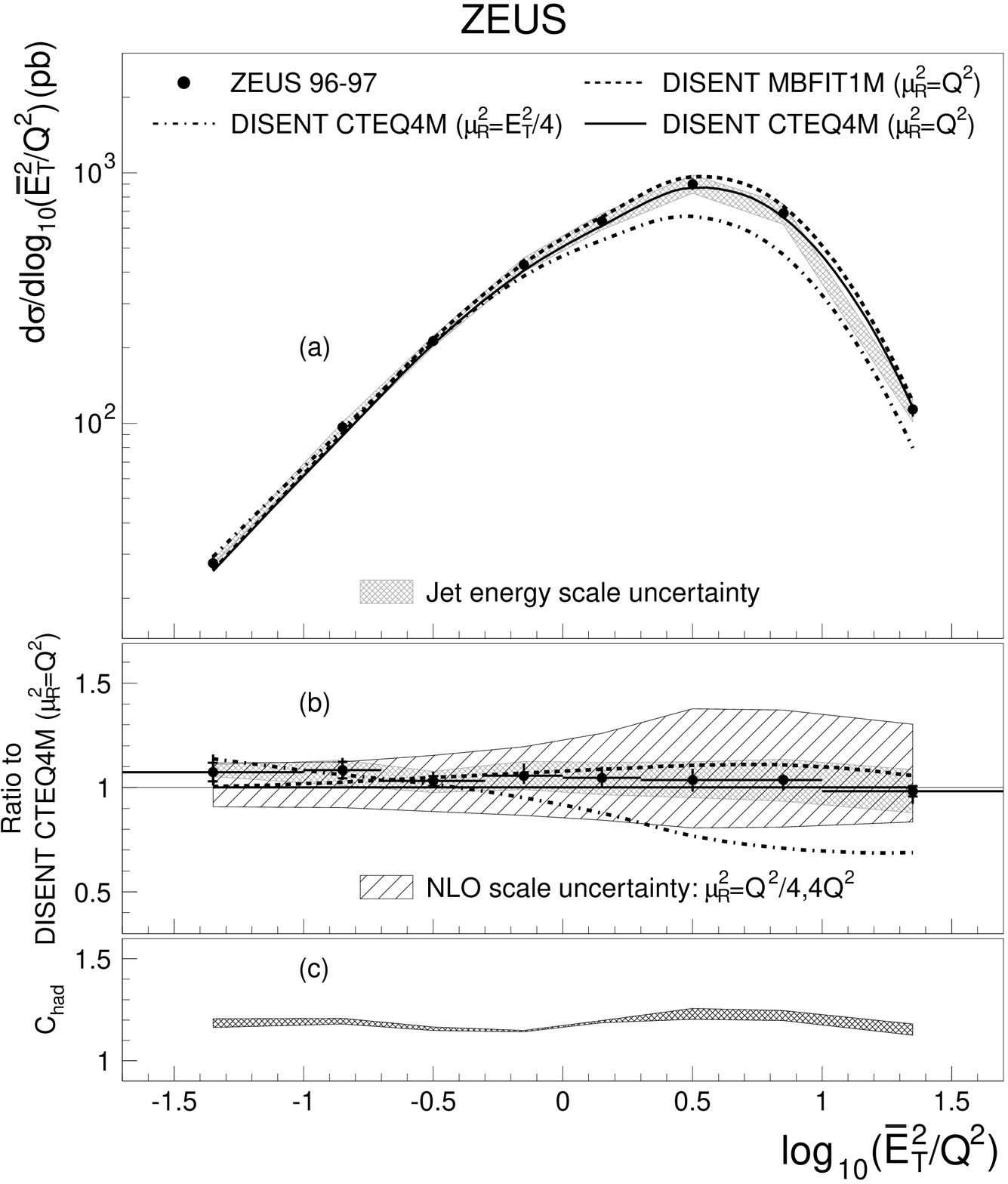,width=18.0cm}
\caption{Dijet cross section, 
  $d\sigma/d\log_{10}(\AVET2/\Q2)$, 
  for jets of hadrons in the Breit frame selected with the inclusive $\kt$ algorithm.
  Other details are as described in the caption of
  Fig.~\ref{fig:q2_cross}.}
\label{fig:aetq}
\end{center}
\end{figure}

The measured dijet cross sections for jets of hadrons as a function of $\log_{10}Q^2$, 
$\log_{10}(\AVET2/\Q2)$ and $\log_{10}\xi$ are given in Tables~\ref{table1}--\ref{table3} and are shown in 
Figs.~\ref{fig:had_sigmas}--\ref{fig:xgluon}. In Fig.~\ref{fig:had_sigmas}, the predictions of ARIADNE and 
LEPTO are compared to the data.  The MC models are in general agreement with the shapes of the 
distributions but fail to describe the normalisation of the measured cross sections. Both models are LO 
and thus the uncertainties due to the choice of $\mu_R^2$ are substantial.

The inclusive dijet differential cross section, $d\sigma/d\log_{10}\Q2$, for jets of hadrons is compared 
to the predictions of NLO QCD in Fig.~\ref{fig:q2_cross}(a). The measured differential cross section 
decreases by two orders of magnitude in the range 10 $<Q^2$ $<$ 10$^4$ GeV$^2$ and is well described by 
the predictions using $\mu_R^2=Q^2$. To examine deviations between the measurement and the predictions, 
the ratio of the measured cross sections and the predictions of DISENT using $\mu_R^2=Q^2$ and CTEQ4M is 
shown in Fig.~\ref{fig:q2_cross}(b). The NLO predictions are consistent with the measurements to within 
10$\%$ for the entire $\Q2$ range. This difference is attributed to the fact that the measured 
cross section includes events with two or more jets, whereas the NLO QCD calculation is limited  to final 
states with two or three partons only\footnote{This hypothesis is supported 
by the fact that the exclusive dijet cross section (for two and only two jets)
at high $\Q2$ has been shown~\cite{Etassi} to agree to within 1$\%$ with the predictions of DISENT using $\mu_R^2 = \Q2$.}. 
The study of inclusive dijets in this paper will permit comparison to higher-order calculations.

The measured cross sections were also compared to the predictions of DISENT using $\mu_R^2~=~E_T^2/4$. For 
$\Q2 > $ 400 GeV$^2$, no significant difference is seen between the predictions with 
$\mu_R^2=Q^2$ and with $\mu_R^2=E_T^2/4$. For lower values of $\Q2$, the predictions with 
$\mu_R^2=E_T^2/4$ underestimate the measured cross sections by as much as 30$\%$. Figure~\ref{fig:q2_cross} 
also shows that the predictions of DISENT with $\mu_R^2=Q^2$ using the proton PDFs from MBFIT1M~\cite{Mbotje} 
and those using the CTEQ4M PDFs agree equally well with the data. The NLO QCD calculations exhibit a significant 
uncertainty due to the choice of $\mu_R^2$. The scale uncertainty  decreases from $^{+50}_{-20}\%$ at 
$\Q2 \approx$ 15 GeV$^2$ to $\pm10\%$ at $\Q2=$~400~GeV$^2$ and $\pm5\%$ at $\Q2=10^4$~GeV$^2$. Only for 
$\Q2 >$ 200 GeV$^2$ are the theoretical uncertainties comparable to the systematic uncertainties of the data. 

To examine the interplay of the $Q$ and $E_T$ scales, 
$d\sigma/d\log_{10}(\AVET2/\Q2)$ is compared to the predictions of DISENT in Fig.~\ref{fig:aetq}. The 
predictions with $\mu^2_R=\Q2$ using either CTEQ4M or MBFIT1M agree well with the data. The scale uncertainty 
is $\pm (15-20)\%$ in the region where $\AVET2 < \Q2$ and rises to $^{+40}_{-20}\%$ for $\AVET2 > \Q2$. It 
is noteworthy that the prediction with $\mu_R^2=Q^2$ gives an excellent description of the data for all 
values of $\AVET2/\Q2$, whereas, as can be seen from Fig.~\ref{fig:aetq}(b), the predictions using 
$\mu_R^2=E_T^2/4$ fail to describe the data for $\bar{E_T}^2 > Q^2$.

\subsection{Gluon distribution of the proton}

\begin{figure}[htp] 
\begin{center}
  \epsfig{figure=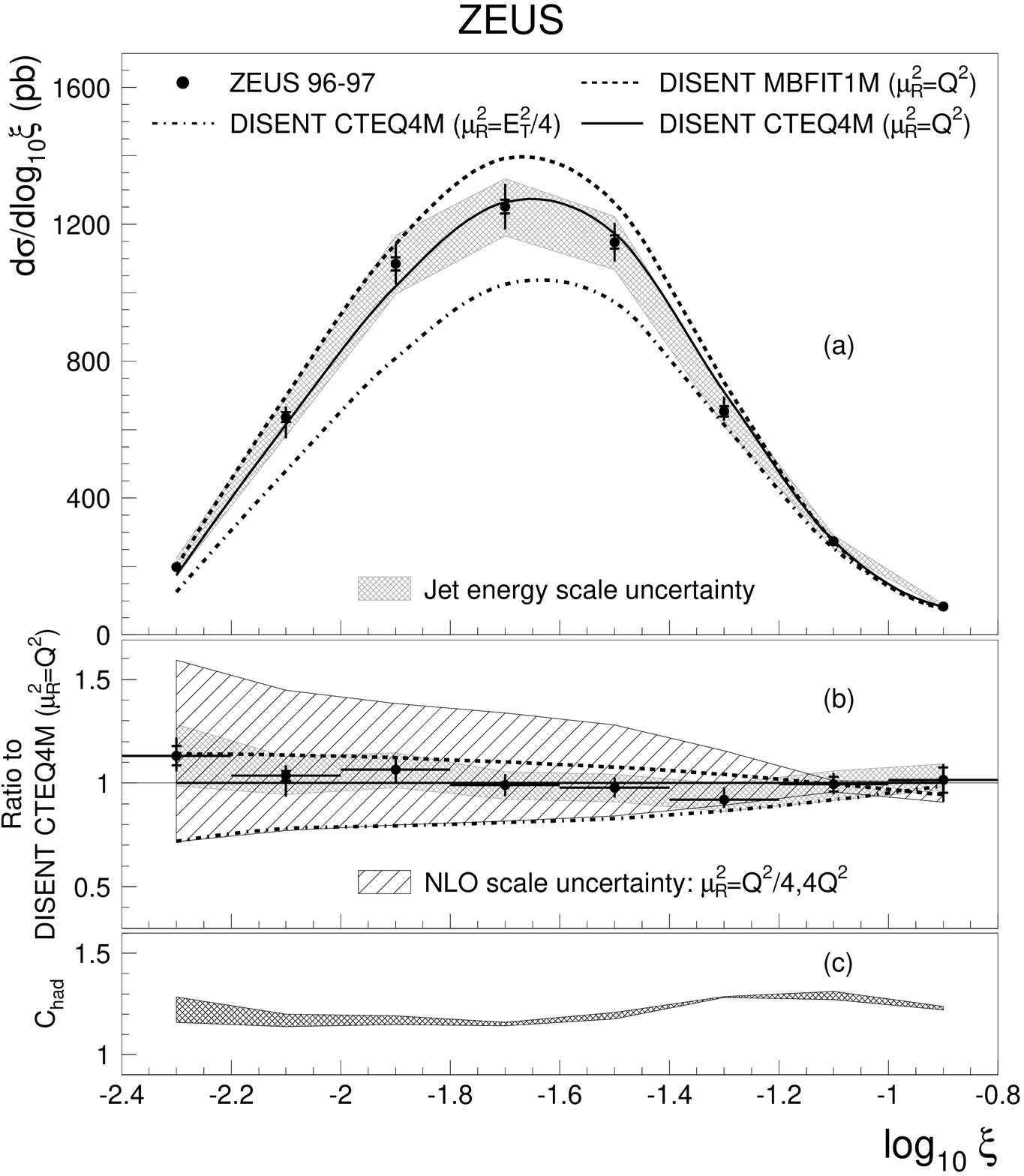,width=14.cm}
  \caption{Dijet cross section, $d\sigma/d\log_{10}\xi$, 
    for jets of hadrons in the Breit frame selected with the inclusive $\kt$ algorithm.
    Other details are as described in the caption of
    Fig.~\ref{fig:q2_cross}.}
  \label{fig:xgluon}
\end{center}
\end{figure}

Figure~\ref{fig:xgluon}(a) shows $d\sigma/d\log_{10}\xi$. 
The requirement that two jets with high transverse energy be observed in the final state suppresses the 
cross section in the low-$\xi$ region. Therefore, the measured dijet cross section rises in the interval 
$-2.3 < \log_{10}\xi < -1.7$. For $\log_{10}\xi > -1.7$, the cross section decreases due to the decrease 
of the gluon and quark densities at high $\xbj$. 

In Fig.~\ref{fig:xgluon}(a), the measured cross sections are compared to the predictions of NLO QCD 
calculated with DISENT using both $\mu_R^2=Q^2$ and $\mu_R^2=E_T^2/4$. The predictions with $\mu_R^2=Q^2$ 
are shown for both the CTEQ4M and the MBFIT1M proton PDFs. Figure~\ref{fig:xgluon}(b) demonstrates that 
the NLO QCD predictions have large theoretical uncertainties over most of the $\xi$ region. These 
uncertainties arise from the sensitivity to the choice of $\mu_R^2$ in the low-$\Q2$ region, as shown in 
Fig.~\ref{fig:q2_cross}(b). 
Given these uncertainties, all of the pQCD calculations shown are consistent with the data.

The measured dijet differential cross sections as a function of $\xi$ for different $\Q2$ ranges are given 
in Tables~\ref{table4} and \ref{table5} and compared to the NLO QCD predictions in Figs.~\ref{fig:double1} 
and~\ref{fig:double2}. At low $\Q2$, the uncertainty on the measurements is similar to the difference 
between the predictions using CTEQ4M and MBFIT1M. As $\Q2$ increases, the dijet cross section is limited by 
the kinematic and jet cuts to high $\xi$ values. 
For $\Q2 >$ 215 GeV$^2$, all of the predictions of NLO QCD 
give consistent results to $\pm 5\%$, independent of the choice of $\mu_R^2$ ($\SET2$/4 or $\Q2$) or proton 
PDFs (CTEQ4M or MBFIT1M). The NLO renormalisation-scale uncertainty in this range decreases from 10$\%$ to 
5$\%$ as $\Q2$ increases. Over the entire $\Q2$ range of these measurements, the predictions of DISENT with 
$\mu_R^2 = \Q2$ are in reasonable agreement with the data.

In the range $10<\Q2<464$ GeV$^2$, the total experimental 
uncertainty of the measurement ranges from $(10-15)\%$ and the theoretical uncertainty, due to the variation of $\mu_R^2$, 
varies from $45\%$ to $10\%$. 
In the low-$\Q2$ region, the dijet sample is dominated by gluon-induced BGF events and so could, in 
principle, be  used to determine the gluon density. However, the size of the scale uncertainties precludes 
a precise extraction of the gluon density\footnote{This situation is also true for exclusive dijet production.}. The 
agreement between the data and the DISENT calculation indicates that the parametrisations of the gluon 
distribution in CTEQ4M and MBFIT1M, which are determined primarily from the 
measurements of the proton 
structure function, are consistent with these dijet measurements.

\section{Conclusions}

Differential dijet cross sections in neutral current deep inelastic scattering have been measured in the 
ranges 10 $< Q^2 <$ 10$^4$ GeV$^2$, $y > $ 0.04, $E^\prime_e> 10$ GeV, $\ETBO > 8$ GeV, $\ETBT > 5$ GeV, 
$\ETLO > 5$ GeV, $\ETLT > 5$ GeV and $|\eta^{\rm LAB}_{1,2}| < 2$ using the longitudinally invariant $\kt$ 
cluster algorithm in the Breit frame. 

The measured cross sections were compared to next-to-leading-order QCD calculations as implemented in 
DISENT and MEPJET. The measured dijet cross sections fall by two orders of magnitude over the $\Q2$ range 
considered here. The NLO QCD predictions agree with the data to within 10$\%$ when the renormalisation scale 
$\mu_R$ is chosen to be $Q$. The estimated theoretical uncertainties of up to $50\%$ for $\Q2<20$ GeV$^2$ 
arise from the absence of higher-order terms, leading to a sensitivity on the choice of the renormalisation 
scale. Only for $\Q2 >$ 200 GeV$^2$ are the theoretical uncertainties comparable to the systematic 
uncertainties of the data. The NLO predictions agree with the measured $\AVET2/\Q2$ and $\xi$ distributions 
to within 10$\%$. Again, for these distributions, the renormalisation-scale uncertainty is large; up to 
40$\%$ when $\AVET2>\Q2$ and 50$\%$ when $\log_{10}\xi<-2$.

At low $\Q2$, the uncertainty on the measurements is similar to the difference between the 
predictions using the CTEQ4M and MBFIT1M parametrisations of the proton PDFs. In the range 
$10<\Q2<464$~GeV$^2$, the total experimental uncertainty of the measurement ranges from $(10-15)\%$ and the 
theoretical uncertainty, due to the variation of $\mu_R^2$, varies from $45\%$ to $10\%$. Within these 
uncertainties, the measurements are consistent with a QCD prediction based on a gluon distribution extracted 
from scaling violations of the $F_2$ structure function of the proton. In this sense, therefore, these 
measurements are consistent with a universal gluon PDF. However, the large theoretical uncertainty precludes 
a useful determination of the gluon PDF in the proton using the dijet
data; improved calculations are needed to exploit the full 
potential of this data. 

\vspace{0.5cm}

\noindent {\Large\bf Acknowledgements}

\vspace{0.3cm}

The design, construction and installation of the ZEUS detector have
been made possible by the ingenuity and dedicated efforts of many
people from inside DESY and from the home institutes who are not
listed as authors. Their contributions are acknowledged with great
appreciation. The experiment was made possible by the inventiveness
and the diligent efforts of the HERA machine group.  The strong
support and encouragement of the DESY directorate have been
invaluable. We would like to thank M. H. Seymour and D. Zeppenfeld 
for useful discussions.

\setcounter{secnumdepth}{0} 



\newpage
\clearpage


\renewcommand{\arraystretch}{1.5}

\begin{table}
\begin{center}
\begin{tabular}[b]{|l|cccc|l|}
\hline

\multicolumn{1}{|c}{log$_{10} \Q2$ bin} & 
\multicolumn{1}{|c}{$d\sigma/d\log_{10}Q^2$} & 
\multicolumn{1}{c}{$\Delta_{\rm stat}$} &
\multicolumn{1}{c}{$\Delta_{\rm sys}$} & 
\multicolumn{1}{c}{$\Delta_{\rm ES}$ (pb)} & 
\multicolumn{1}{|c|}{ISR/FSR} \\

\multicolumn{1}{|c}{} & 
\multicolumn{1}{|c}{} & 
\multicolumn{1}{c}{} & 
\multicolumn{1}{c}{} & 
\multicolumn{1}{c}{} & 
\multicolumn{1}{|c|}{correction} \\

\hline
1.00, 1.20 &  1120.5 & $\pm$ 18.5 & $  ^{+154.3}_{-141.4}$ & $ ^{+92.3}_{-93.0} $ & $ 1.024 \pm 0.012 $ \\
1.20, 1.40 &  967.7  & $\pm$ 17.7 & $ ^{+80.0}_{-95.8}  $  & $ ^{+84.5}_{-85.6} $ & $ 1.040 \pm 0.014 $ \\
1.40, 1.60 &  853.6  & $\pm$ 16.7 & $ ^{+46.0}_{-66.9}  $  & $ ^{+68.5}_{-75.0} $ & $ 0.990 \pm 0.014 $ \\
1.60, 1.80 &  690.5  & $\pm$ 15.1 & $ ^{+34.3}_{-24.0}  $  & $ ^{+60.9}_{-56.3} $ & $ 1.002 \pm 0.016 $ \\
1.80, 2.00 &  549.4  & $\pm$ 13.3 & $ ^{+35.0}_{-14.6}  $  & $ ^{+45.5}_{-44.2} $ & $ 0.987 \pm 0.018 $ \\
2.00, 2.33 &  368.2  & $\pm$ 7.3  & $ ^{+16.1}_{-11.5}  $  & $ ^{+24.1}_{-26.1} $ & $ 0.989 \pm 0.012 $ \\
2.33, 2.67 &  219.1  & $\pm$ 5.1  & $ ^{+5.3}_{-5.3}    $  & $ ^{+13.8}_{-14.0} $ & $ 1.004 \pm 0.011 $ \\
2.67, 3.00 &  113.5  &  $\pm$ 3.5  & $ ^{+2.2}_{-4.3}    $  & $ ^{+5.7}_{-5.9} $ & $ 1.021 \pm 0.012 $ \\
3.00, 3.50 &  44.2   &  $\pm$ 1.6  & $ ^{+2.4}_{-0.6}    $  & $ ^{+2.2}_{-2.4} $ & $ 0.980 \pm 0.007 $ \\
3.50, 4.00 &  8.7    &  $\pm$ 0.7  & $ ^{+0.8}_{-0.7}    $  & $ ^{+0.3}_{-0.4} $ & $ 0.960 \pm 0.008 $ \\
\hline
\end{tabular}
\caption{ \label{table1} Dijet cross sections, $d\sigma/d\log_{10}Q^2$, where $Q^2$ is in GeV$^2$, for jets of 
hadrons in the Breit frame, selected with the inclusive $\kt$ algorithm in the ranges 10 $< Q^2 <$ 10$^4$ GeV$^2$, 
$y > $ 0.04, $E^\prime_e>~10$ GeV, $\ETBO > 8$ GeV, $\ETBT > 5$ GeV, $\ETLO > 5$ GeV, $\ETLT > 5$ GeV 
and $|\eta^{\rm LAB}_{1,2}| < 2$. The statistical, systematic and jet-energy-scale uncertainties are shown 
separately. The multiplicative correction applied to account for effects of initial- and final-state QED 
radiation is shown in the last column.}
\end{center}
\end{table}

\newpage
\clearpage


\begin{table}
\begin{center}
\begin{tabular}[b]{|c|cccc|l|}
\hline
\multicolumn{1}{|c}{$\log_{10}(\AVET2/\Q2)$ bin} & 
\multicolumn{1}{|c}{$d\sigma/d\log_{10}(\AVET2/\Q2)$} & 
\multicolumn{1}{c}{$\Delta_{\rm stat}$} &
\multicolumn{1}{c}{$\Delta_{\rm sys}$} & 
\multicolumn{1}{c}{$\Delta_{\rm ES}$ (pb)} & 
\multicolumn{1}{|c|}{ISR/FSR} \\

\multicolumn{1}{|c}{} & 
\multicolumn{1}{|c}{} & 
\multicolumn{1}{c}{} & 
\multicolumn{1}{c}{} & 
\multicolumn{1}{c}{} & 
\multicolumn{1}{|c|}{correction} \\

\hline
$-$1.70, $-$1.00 &  27.6  & $\pm$ 1.1  & $^{+1.8}_{-1.2}  $ & $ ^{+0.9}_{-0.6}   $ & $ 0.943 \pm 0.010 $ \\
$-$1.00, $-$0.70 &  96.5  & $\pm$ 3.4  & $^{+3.9}_{-2.4}  $ & $ ^{+4.3}_{-6.7}   $ & $ 0.978 \pm 0.013 $ \\
$-$0.70, $-$0.30 &  213.2 & $\pm$ 5.0  & $^{+7.5}_{-5.2}  $ & $ ^{+9.4}_{-11.4}  $ & $ 1.009 \pm 0.012 $ \\
$-$0.30, +0.00 &  429.5 & $\pm$ 9.4  & $^{+21.6}_{-12.6}$ & $ ^{+27.9}_{-26.6} $ & $ 1.000 \pm 0.014 $ \\
+0.00, +0.30 &  640.9 & $\pm$ 12.2 & $^{+30.7}_{-26.7}$ & $ ^{+46.3}_{-48.8} $ & $ 0.994 \pm 0.013 $ \\
+0.30, +0.70 &  901.3 & $\pm$ 12.7 & $^{+45.0}_{-45.6}$ & $ ^{+75.7}_{-73.5} $ & $ 1.010 \pm 0.010 $ \\
+0.70, +1.00 &  688.7 & $\pm$ 12.5 & $^{+14.6}_{-30.4}$ & $ ^{+64.7}_{-67.4} $ & $ 1.021 \pm 0.013 $ \\
+1.00, +1.70 &  113.6 & $\pm$ 2.8  & $^{+1.9}_{-6.4}  $ & $ ^{+11.9}_{-11.9} $ & $ 1.029 \pm 0.017 $ \\
\hline
\end{tabular}
\caption{\label{table2} Dijet cross sections, $d\sigma/d\log_{10}(\AVET2/\Q2)$, 
for jets of hadrons in the Breit frame, selected with the inclusive $\kt$ 
algorithm. For details, see the caption of Table 1.}
\end{center}
\end{table}

\newpage
\clearpage


\begin{table}
\begin{center}
\begin{tabular}[b]{|l|cccc|l|}
\hline
\multicolumn{1}{|c}{$\log_{10}\xi$ bin} & 
\multicolumn{1}{|c}{$d\sigma/d\log_{10}\xi$} & 
\multicolumn{1}{c}{$\Delta_{\rm stat}$} &
\multicolumn{1}{c}{$\Delta_{\rm sys}$} & 
\multicolumn{1}{c}{$\Delta_{\rm ES}$} & 
\multicolumn{1}{|c|}{ISR/FSR} \\

\multicolumn{1}{|c}{} & 
\multicolumn{1}{|c}{} & 
\multicolumn{1}{c}{} & 
\multicolumn{1}{c}{} & 
\multicolumn{1}{c}{} & 
\multicolumn{1}{|c|}{correction} \\

\hline
$-$2.40, $-$2.20 & 198.2  & $\pm$ 8.1  & $^{+13.1}_{-10.4}$ & $ ^{+26.4}_{-25.6} $ & $ 1.138 \pm 0.037 $ \\
$-$2.20, $-$2.00 & 636.4  & $\pm$ 14.8 & $^{+27.7}_{-59.8}$ & $ ^{+55.8}_{-56.1} $ & $ 1.052 \pm 0.018 $ \\
$-$2.00, $-$1.80 & 1084.5 & $\pm$ 19.5 & $^{+57.7}_{-57.2}$ & $ ^{+83.0}_{-88.4} $ & $ 1.043 \pm 0.013 $ \\
$-$1.80, $-$1.60 & 1251.4 & $\pm$ 20.6 & $^{+63.3}_{-63.1}$ & $ ^{+81.7}_{-85.0} $ & $ 0.995 \pm 0.011 $ \\
$-$1.60, $-$1.40 & 1147.9 & $\pm$ 20.0 & $^{+52.7}_{-53.6}$ & $ ^{+74.2}_{-80.2} $ & $ 0.980 \pm 0.011 $ \\
$-$1.40, $-$1.20 & 653.7  & $\pm$ 15.0 & $^{+39.8}_{-24.1}$ & $ ^{+53.4}_{-42.4} $ & $ 0.994 \pm 0.013 $ \\
$-$1.20, $-$1.00 & 273.9  & $\pm$ 9.5  & $^{+11.1}_{-9.7} $ & $ ^{+17.5}_{-21.5} $ & $ 0.986 \pm 0.019 $ \\
$-$1.00, $-$0.80 & 82.6   & $\pm$ 5.0  & $^{+3.8}_{-7.0}  $ & $ ^{+6.3}_{-3.9}   $ & $ 0.997 \pm 0.031 $ \\

\hline
\end{tabular}
\caption{\label{table3} Dijet cross sections, $d\sigma/d\log_{10}\xi$, for jets of hadrons in the Breit 
frame, selected with the inclusive $\kt$ algorithm. For details, see the caption of Table 1.}
\end{center}
\end{table}

\renewcommand{\arraystretch}{1.2}

\newpage
\clearpage
\begin{table}
\begin{center}
\begin{tabular}[b]{|l|cccc|l|}
\hline
\multicolumn{1}{|c}{$\log_{10}\xi$ bin} & 
\multicolumn{1}{|c}{$d\sigma/d\log_{10}\xi$} & 
\multicolumn{1}{c}{$\Delta_{\rm stat}$} &
\multicolumn{1}{c}{$\Delta_{\rm sys}$} & 
\multicolumn{1}{c}{$\Delta_{\rm ES}$ (pb)} & 
\multicolumn{1}{|c|}{ISR/FSR} \\

\multicolumn{1}{|c}{} & 
\multicolumn{1}{|c}{} & 
\multicolumn{1}{c}{} & 
\multicolumn{1}{c}{} & 
\multicolumn{1}{c}{} & 
\multicolumn{1}{|c|}{correction} \\

\hline
\hline
\multicolumn{6}{|c|}{$1.0 < \log_{10}Q^{2} < 1.2$} \\
\hline
$-$2.40, $-$2.20 & $ 70.2  $ & $ \pm 4.9 $ & $ ^{+7.0}_{-3.9}   $ & $ ^{+6.6}_{-7.2}   $ & $ 1.173 \pm 0.065 $ \\
$-$2.20, $-$2.00 & $ 169.9 $ & $ \pm 7.7 $ & $ ^{+20.3}_{-21.5} $ & $ ^{+14.6}_{-16.7} $ & $ 1.021 \pm 0.034 $ \\
$-$2.00, $-$1.80 & $ 261.9 $ & $ \pm 9.9 $ & $ ^{+25.9}_{-27.8} $ & $ ^{+21.2}_{-19.9} $ & $ 1.073 \pm 0.028 $ \\
$-$1.80, $-$1.60 & $ 251.5 $ & $ \pm 9.3 $ & $ ^{+22.2}_{-21.4} $ & $ ^{+15.7}_{-15.4} $ & $ 1.002 \pm 0.024 $ \\
$-$1.60, $-$1.40 & $ 199.6 $ & $ \pm 8.4 $ & $ ^{+18.8}_{-21.7} $ & $ ^{+13.9}_{-16.9} $ & $ 1.012 \pm 0.026 $ \\
$-$1.40, $-$1.20 & $ 78.4  $ & $ \pm 5.0 $ & $ ^{+12.9}_{-3.7}  $ & $ ^{+6.3}_{-5.2}   $ & $ 0.967 \pm 0.034 $ \\
\hline
\multicolumn{6}{|c|}{$1.2 < \log_{10}Q^{2} < 1.4$} \\
\hline
$-$2.40, $-$2.20 & $ 55.9  $ & $ \pm 4.6 $ & $ ^{+1.5}_{-7.1}   $ & $ ^{+8.1}_{-8.8}   $ & $ 1.214 \pm 0.077 $ \\
$-$2.20, $-$2.00 & $ 142.2 $ & $ \pm 7.2 $ & $ ^{+13.2}_{-13.5} $ & $ ^{+14.8}_{-11.3} $ & $ 1.092 \pm 0.041 $ \\
$-$2.00, $-$1.80 & $ 222.5 $ & $ \pm 9.3 $ & $ ^{+7.3}_{-13.3}  $ & $ ^{+17.8}_{-19.6} $ & $ 1.089 \pm 0.032 $ \\
$-$1.80, $-$1.60 & $ 212.8 $ & $ \pm 8.6 $ & $ ^{+17.0}_{-10.7} $ & $ ^{+14.9}_{-10.8} $ & $ 1.003 \pm 0.027 $ \\
$-$1.60, $-$1.40 & $ 181.2 $ & $ \pm 8.2 $ & $ ^{+9.4}_{-14.9}  $ & $ ^{+11.2}_{-15.1} $ & $ 0.968 \pm 0.028 $ \\
$-$1.40, $-$1.20 & $ 83.3  $ & $ \pm 5.6 $ & $ ^{+15.0}_{-2.2}  $ & $ ^{+6.1}_{-7.1}   $ & $ 1.108 \pm 0.042 $ \\
\hline
\multicolumn{6}{|c|}{$1.4 < \log_{10}Q^{2} < 1.6$} \\
\hline
$-$2.40, $-$2.20 & $ 33.9  $ & $ \pm 3.2 $ & $ ^{+3.8}_{-2.4}  $ & $ ^{+5.9}_{-3.2}   $ & $ 1.084 \pm 0.083 $ \\
$-$2.20, $-$2.00 & $ 128.0 $ & $ \pm 7.0 $ & $ ^{+3.0}_{-17.2} $ & $ ^{+8.2}_{-11.3}  $ & $ 1.064 \pm 0.044 $ \\
$-$2.00, $-$1.80 & $ 182.7 $ & $ \pm 8.2 $ & $ ^{+2.9}_{-8.7}  $ & $ ^{+12.9}_{-16.9} $ & $ 0.988 \pm 0.032 $ \\
$-$1.80, $-$1.60 & $ 198.4 $ & $ \pm 8.5 $ & $ ^{+8.8}_{-19.2} $ & $ ^{+12.6}_{-16.6} $ & $ 0.938 \pm 0.028 $ \\
$-$1.60, $-$1.40 & $ 165.2 $ & $ \pm 7.9 $ & $ ^{+11.7}_{-8.6} $ & $ ^{+13.5}_{-9.6}  $ & $ 1.000 \pm 0.031 $ \\
$-$1.40, $-$1.20 & $ 89.5  $ & $ \pm 6.1 $ & $ ^{+5.1}_{-6.7}  $ & $ ^{+7.9}_{-9.5}   $ & $ 0.977 \pm 0.040 $ \\
\hline
\multicolumn{6}{|c|}{$1.6 < \log_{10}Q^{2} < 1.8$} \\
\hline
$-$2.40, $-$2.20 & $ 25.5  $ & $ \pm 2.9 $ & $ ^{+1.2}_{-1.4}  $ & $ ^{+3.2}_{-3.9}   $ & $ 1.085 \pm 0.100 $ \\
$-$2.20, $-$2.00 & $ 83.0  $ & $ \pm 5.3 $ & $ ^{+1.7}_{-7.7}  $ & $ ^{+8.7}_{-7.6}   $ & $ 1.032 \pm 0.050 $ \\
$-$2.00, $-$1.80 & $ 146.4 $ & $ \pm 7.3 $ & $ ^{+11.6}_{-8.8} $ & $ ^{+11.5}_{-12.1} $ & $ 1.059 \pm 0.038 $ \\
$-$1.80, $-$1.60 & $ 172.0 $ & $ \pm 8.0 $ & $ ^{+21.3}_{-3.9} $ & $ ^{+13.2}_{-11.2} $ & $ 1.000 \pm 0.033 $ \\
$-$1.60, $-$1.40 & $ 149.1 $ & $ \pm 7.6 $ & $ ^{+3.8}_{-1.7}  $ & $ ^{+9.7}_{-10.8}  $ & $ 0.942 \pm 0.032 $ \\
$-$1.40, $-$1.20 & $ 69.2  $ & $ \pm 5.2 $ & $ ^{+3.5}_{-6.0}  $ & $ ^{+8.4}_{-3.9}   $ & $ 0.988 \pm 0.046 $ \\
\hline
\multicolumn{6}{|c|}{$1.8 < \log_{10}Q^{2} < 2.0$} \\
\hline
$-$2.20, $-$2.00 & $ 61.6  $ & $ \pm 4.5 $ & $ ^{+2.6}_{-5.7} $ & $ ^{+6.5}_{-4.7}  $ & $ 1.019 \pm 0.057 $ \\
$-$2.00, $-$1.80 & $ 113.0 $ & $ \pm 6.3 $ & $ ^{+4.5}_{-3.7} $ & $ ^{+6.2}_{-8.6}  $ & $ 1.008 \pm 0.041 $ \\
$-$1.80, $-$1.60 & $ 133.7 $ & $ \pm 6.9 $ & $ ^{+8.7}_{-3.3} $ & $ ^{+12.2}_{-9.6} $ & $ 1.019 \pm 0.037 $ \\
$-$1.60, $-$1.40 & $ 124.5 $ & $ \pm 6.8 $ & $ ^{+7.0}_{-3.7} $ & $ ^{+8.0}_{-9.6}  $ & $ 0.941 \pm 0.035 $ \\
$-$1.40, $-$1.20 & $ 66.8  $ & $ \pm 5.1 $ & $ ^{+7.2}_{-5.2} $ & $ ^{+8.1}_{-5.2}  $ & $ 0.943 \pm 0.044 $ \\
\hline
\end{tabular}
\caption{Dijet cross sections, $d\sigma/d\log_{10}\xi$, for jets of hadrons in the Breit frame, selected 
with the inclusive $\kt$ algorithm. For details, see the caption of Table 1.}
\label{table4}
\end{center}
\end{table}

\newpage
\clearpage
\begin{table}
\begin{center}
\begin{tabular}[b]{|l|cccc|l|}
\hline
\multicolumn{1}{|c}{$\log_{10}\xi$ bin} & 
\multicolumn{1}{|c}{$d\sigma/d\log_{10}\xi$} & 
\multicolumn{1}{c}{$\Delta_{\rm stat}$} &
\multicolumn{1}{c}{$\Delta_{\rm sys}$} & 
\multicolumn{1}{c}{$\Delta_{\rm ES}$} & 
\multicolumn{1}{|c|}{ISR/FSR} \\

\multicolumn{1}{|c}{} & 
\multicolumn{1}{|c}{} & 
\multicolumn{1}{c}{} & 
\multicolumn{1}{c}{} & 
\multicolumn{1}{c}{} & 
\multicolumn{1}{|c|}{correction} \\
\hline
\hline
\multicolumn{6}{|c|}{$2.0 < \log_{10}Q^{2} < 2.33$} \\
\hline
$-$2.20, $-$2.00 & $ 43.2  $ & $ \pm 3.2 $ & $ ^{+5.3}_{-1.7} $ & $ ^{+2.3}_{-3.2}  $ & $ 1.126 \pm 0.054 $ \\
$-$2.00, $-$1.80 & $ 105.9 $ & $ \pm 5.1 $ & $ ^{+9.5}_{-7.0} $ & $ ^{+8.3}_{-8.5}  $ & $ 0.997 \pm 0.029 $ \\
$-$1.80, $-$1.60 & $ 153.2 $ & $ \pm 6.5 $ & $ ^{+3.6}_{-7.6} $ & $ ^{+8.1}_{-11.7} $ & $ 0.980 \pm 0.024 $ \\
$-$1.60, $-$1.40 & $ 148.7 $ & $ \pm 6.5 $ & $ ^{+6.4}_{-3.1} $ & $ ^{+9.3}_{-7.1}  $ & $ 0.972 \pm 0.023 $ \\
$-$1.40, $-$1.20 & $ 94.7  $ & $ \pm 5.4 $ & $ ^{+2.2}_{-6.6} $ & $ ^{+6.3}_{-5.8}  $ & $ 0.981 \pm 0.028 $ \\
\hline
\multicolumn{6}{|c|}{$2.33 < \log_{10}Q^{2} < 2.67$} \\
\hline
$-$2.00, $-$1.80 & $ 44.2  $ & $ \pm 3.0 $ & $ ^{+2.3}_{-3.3} $ & $ ^{+4.6}_{-2.3} $ & $ 1.030 \pm 0.032 $ \\
$-$1.80, $-$1.60 & $ 87.1  $ & $ \pm 4.5 $ & $ ^{+3.0}_{-1.9} $ & $ ^{+2.5}_{-5.6} $ & $ 1.062 \pm 0.024 $ \\
$-$1.60, $-$1.40 & $ 104.0 $ & $ \pm 5.1 $ & $ ^{+3.9}_{-5.2} $ & $ ^{+5.2}_{-7.6} $ & $ 0.988 \pm 0.020 $ \\
$-$1.40, $-$1.20 & $ 75.7  $ & $ \pm 4.2 $ & $ ^{+7.6}_{-2.4} $ & $ ^{+5.9}_{-2.5} $ & $ 0.964 \pm 0.021 $ \\
$-$1.20, $-$1.00 & $ 38.1  $ & $ \pm 3.2 $ & $ ^{+3.4}_{-0.8} $ & $ ^{+2.4}_{-3.5} $ & $ 0.979 \pm 0.030 $ \\
\hline
\multicolumn{6}{|c|}{$2.67 < \log_{10}Q^{2} < 3.0$} \\
\hline
$-$2.00, $-$1.80 & $ 6.4  $ & $ \pm 0.9 $ & $ ^{+1.0}_{-0.6} $ & $ ^{+0.6}_{-0.4} $ & $ 1.052 \pm 0.062 $ \\
$-$1.80, $-$1.60 & $ 31.1 $ & $ \pm 2.2 $ & $ ^{+0.4}_{-1.2} $ & $ ^{+1.4}_{-2.5} $ & $ 0.992 \pm 0.029 $ \\
$-$1.60, $-$1.40 & $ 47.5 $ & $ \pm 2.9 $ & $ ^{+3.1}_{-1.9} $ & $ ^{+2.6}_{-2.1} $ & $ 1.010 \pm 0.024 $ \\
$-$1.40, $-$1.20 & $ 54.2 $ & $ \pm 3.3 $ & $ ^{+1.7}_{-4.9} $ & $ ^{+1.6}_{-2.1} $ & $ 1.034 \pm 0.024 $ \\
$-$1.20, $-$1.00 & $ 33.1 $ & $ \pm 2.6 $ & $ ^{+2.7}_{-2.1} $ & $ ^{+2.6}_{-1.9} $ & $ 1.053 \pm 0.029 $ \\
\hline
\multicolumn{6}{|c|}{$3.0 < \log_{10}Q^{2} < 3.5$} \\
\hline
$-$1.80, $-$1.60 & $ 5.2  $ & $ \pm 0.9 $ & $ ^{+1.5}_{-1.5} $ & $ ^{+0.5}_{-0.8} $ & $ 1.060 \pm 0.038 $ \\
$-$1.60, $-$1.40 & $ 18.7 $ & $ \pm 1.7 $ & $ ^{+2.9}_{-1.0} $ & $ ^{+0.6}_{-1.4} $ & $ 0.981 \pm 0.018 $ \\
$-$1.40, $-$1.20 & $ 32.3 $ & $ \pm 2.3 $ & $ ^{+1.3}_{-1.6} $ & $ ^{+2.1}_{-1.5} $ & $ 0.982 \pm 0.014 $ \\
$-$1.20, $-$1.00 & $ 28.9 $ & $ \pm 2.1 $ & $ ^{+2.2}_{-0.6} $ & $ ^{+1.1}_{-0.9} $ & $ 0.968 \pm 0.014 $ \\
$-$1.00, $-$0.80 & $ 18.6 $ & $ \pm 1.7 $ & $ ^{+0.9}_{-1.1} $ & $ ^{+0.4}_{-0.9} $ & $ 0.969 \pm 0.017 $ \\
$-$0.80, $-$0.60 & $ 5.6  $ & $ \pm 0.9 $ & $ ^{+0.7}_{-0.6} $ & $ ^{+0.7}_{-0.3} $ & $ 0.981 \pm 0.032 $ \\
\hline
\multicolumn{6}{|c|}{$3.5 < \log_{10}Q^{2} < 4.0$} \\
\hline
$-$1.40, $-$1.20 & $ 0.9 $ & $ \pm 0.4 $ & $ ^{+0.7}_{-0.0} $ & $ ^{+0.3}_{-0.0} $ & $ 1.092 \pm 0.041 $ \\
$-$1.20, $-$1.00 & $ 6.4 $ & $ \pm 1.0 $ & $ ^{+0.6}_{-1.1} $ & $ ^{+0.0}_{-0.5} $ & $ 0.981 \pm 0.018 $ \\
$-$1.00, $-$0.80 & $ 5.7 $ & $ \pm 0.9 $ & $ ^{+1.0}_{-0.2} $ & $ ^{+0.3}_{-0.1} $ & $ 0.957 \pm 0.014 $ \\
$-$0.80, $-$0.60 & $ 6.9 $ & $ \pm 1.0 $ & $ ^{+0.3}_{-0.7} $ & $ ^{+0.1}_{-0.3} $ & $ 0.923 \pm 0.015 $ \\
$-$0.60, $-$0.40 & $ 1.9 $ & $ \pm 0.5 $ & $ ^{+0.5}_{-0.4} $ & $ ^{+0.1}_{-0.1} $ & $ 0.964 \pm 0.027 $ \\
\hline
\end{tabular}
\caption{Continuation of Table 4. For details,
see the caption of Table 1.}
\label{table5}
\end{center}
\end{table}

\newpage


\newpage


\newpage


\newpage


\newpage

\newpage

\newpage


\newpage


\newpage


\newpage


\begin{figure}[htp]
  \begin{center}
    \epsfig{figure=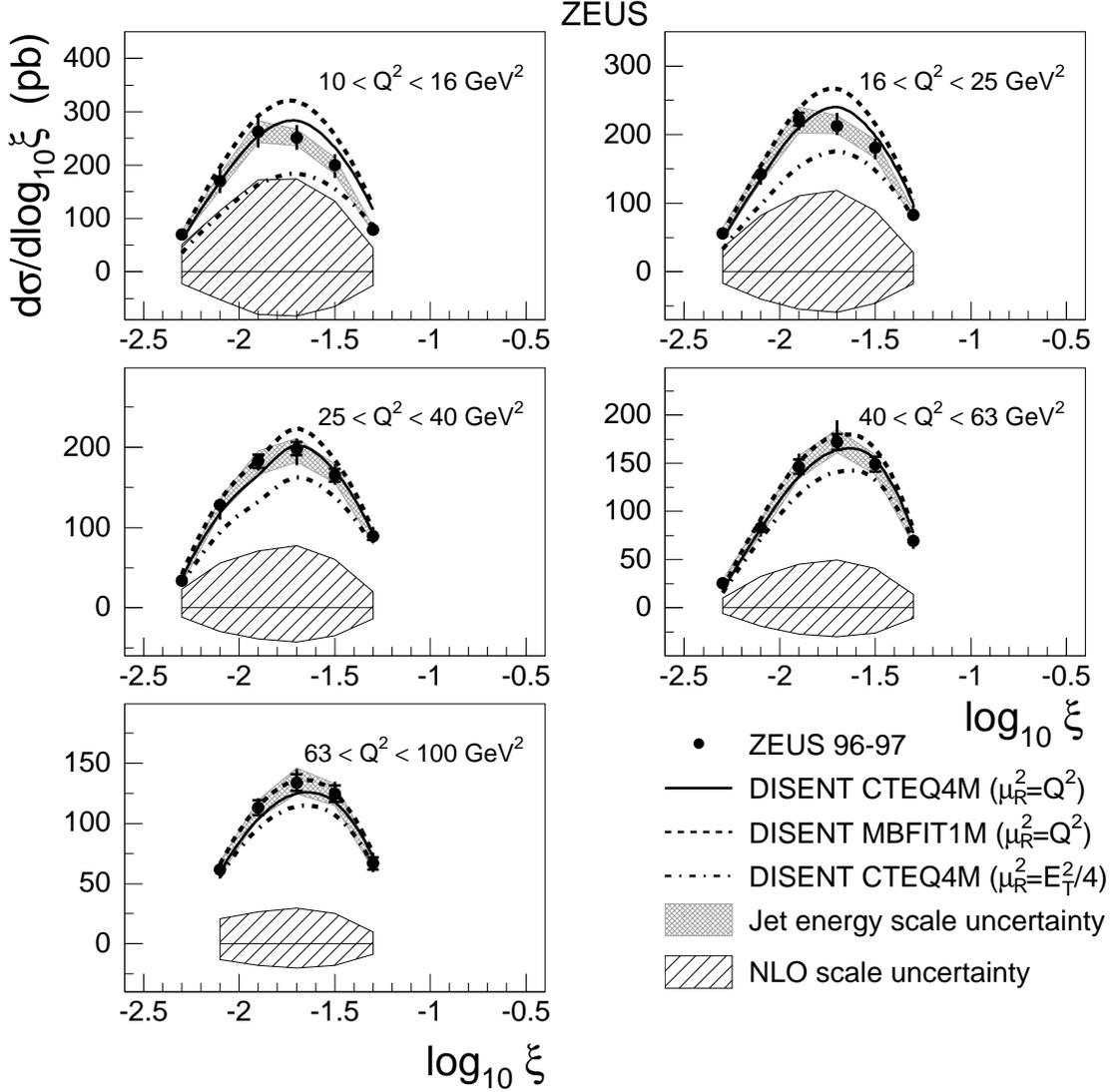,width=16.0cm}\\
    \caption{Dijet cross sections, $d\sigma/d\log_{10}\xi$, for jets of hadrons
      selected with the inclusive $\kt$ jet algorithm in different regions of $\Q2$.  The points
      represent the measured cross sections.  The inner error bars
      represent the statistical uncertainties and the outer bars are the
      statistical and the uncorrelated systematic uncertainties added
      in quadrature. The shaded band represents the systematic
      uncertainty due to the jet-energy scale. The full line
      represents the predictions of NLO QCD using DISENT with
      $\mu_R^2=Q^2$ and the CTEQ4M proton PDFs. The dot-dashed line is
      the prediction of DISENT with $\mu_R^2=E_T^2/4$ and the CTEQ4M
      proton PDFs.  The dashed line is the prediction of DISENT with
      $\mu_R^2=Q^2$ using the MBFIT1M proton PDFs.  The cross-hatched band at
      the bottom of each plot is the $\mu_R^2$ uncertainty calculated
      by varying $\mu_R^2$ in the range (4$\Q2$, $\Q2$/4).  All of the pQCD
      predictions shown here refer to jets of hadrons and were
      obtained by dividing the parton level predictions of DISENT by
      the average value of the correction factors computed using the
      LEPTO~6.5 and the ARIADNE~4.10 MC simulations.}
    \newpage
    \label{fig:double1}
  \end{center}
\end{figure}

\newpage

\begin{figure}[htp]
  \begin{center}
    \epsfig{figure=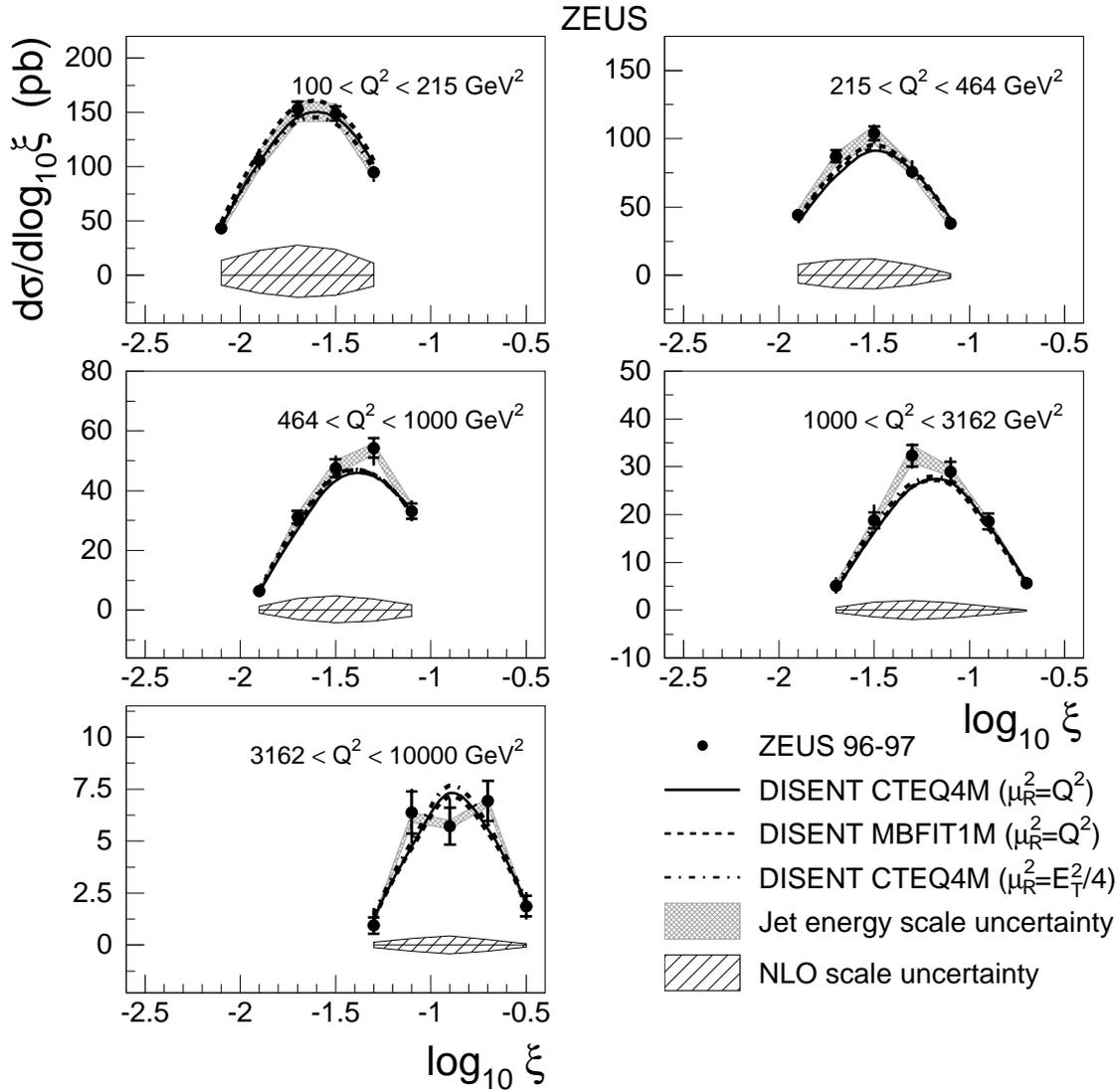,width=16.0cm}
    \caption{Continuation of Fig.~\ref{fig:double1} for higher values of $\Q2$. Details are as described in the caption of Fig.~\ref{fig:double1}.}  
    \label{fig:double2}
  \end{center} 
\end{figure} 

\end{document}